\documentclass[journal,letterpaper,final,10pt]{IEEEtran}

\usepackage{microtype}
\usepackage[vlined, ruled, linesnumbered]{algorithm2e}

\usepackage[utf8x]{inputenc}
\usepackage[caption=false]{subfig}
\usepackage{amsfonts,amsmath,amssymb,amsthm}
\usepackage{float}
\usepackage{graphicx}
\usepackage{epstopdf}
\usepackage{pifont}
\usepackage{multirow}
\usepackage{cite}
\usepackage{array}
\usepackage{cases}
\usepackage{enumitem}
\usepackage{capt-of}
\usepackage{color}
\usepackage{tikz}
\usepackage{bbm}
\usepackage{comment}
\usepackage{url}
\usepackage{soul}

\DeclareCaptionLabelFormat{andtable}%
{#1 ̃#2 \& \tablename ̃\thetable}

\specialcomment{notes}{\begingroup\sffamily\color{black}}{\endgroup}

\newtheorem{thm}{Theorem}
\newtheorem{lm}{Lemma}

\newtheorem{cor}{Corollary}

\newcolumntype{S}{>{\centering\arraybackslash} m{1cm} }
\newcommand{\modf}[1]{\textcolor{black}{#1}} %Christina's comments

\newcommand{\mquote}[1]{``#1''} %Christina's quotes

\title{How CSMA/CA with Deferral Affects Performance and Dynamics in Power-Line Communications \\ \textbf{\large Extended version of our publication in IEEE/ACM Transactions on Networking }}
\author{Christina~Vlachou, Albert~Banchs,~\IEEEmembership{Senior Member, IEEE}, Julien~Herzen and Patrick~Thiran,~\IEEEmembership{Fellow, IEEE}%
\thanks{
	C. Vlachou and P. Thiran are with the \'{E}cole Polytechnique F\'{e}d\'{e}rale de Lausanne, Lausanne 1015, Switzerland (e-mail: christina.vlachou@epfl.ch and patrick.thiran@epfl.ch). A. Banchs is with University Carlos III of Madrid and with IMDEA Networks institute, Legan\'{e}s 28911, Spain (email: banchs@it.uc3m.es). J. Herzen is with Swisscom Bern 3006, Switzerland (email: julien.herzen@swisscom.com).}
\thanks{The work has been financially supported by a grant of the SmartWorld project of the Hasler Foundation, Bern, Switzerland. It was performed during Prof.~Banchs' visit to EPFL. The work of Albert Banchs has been partially funded by the Spanish Ministry MINECO with the THWART project (research grant TEC2015-70836-ERC).
This paper is an extended version of our work published at IEEE/ACM Transactions on Networking.}}

\begin{document}
\maketitle
\begin{abstract}
	Power-line communications are becoming a key component in home networking because they provide easy and high-throughput connectivity. The dominant MAC protocol for high data-rate power-line communications, IEEE 1901, employs a CSMA/CA mechanism similar to the backoff process of 802.11. Existing performance evaluation studies of this protocol assume that the backoff processes of the stations are independent (the so-called \emph{decoupling assumption}). However, in contrast to 802.11, 1901 stations can change their state after sensing the medium busy, which is regulated by the so-called \emph{deferral counter}. This mechanism introduces strong coupling between the stations and, as a result, makes existing analyses inaccurate. In this paper, we propose a performance model for 1901, which does not rely on the decoupling assumption. We prove that our model admits a unique solution for a wide range of configurations and confirm the accuracy of the model using both testbed experiments and simulations. Our results show that we outperform current models based on the decoupling assumption. 
	In addition to evaluating the performance in steady-state, we further study the transient dynamics of 1901, which is also affected by the deferral counter. 
\end{abstract}

\section{Introduction}
\label{sec:intro}
% 1. Global motivation:
\PARstart{P}{ower-line} communications (PLC) are increasingly important in home networking.
HomePlug, the most popular specification for PLC, is employed by over 180 million devices worldwide~\cite{homeplug-nr}, and offers data rates up to 1.5 Gbps.
Moreover, PLC plays an important role in hybrid networks comprising wireless, Ethernet, and other technologies~\cite{hybrid_nets}, as it contributes to increasing the bandwidth of such networks with an independent, widely~accessible~medium.
Yet, despite the wide adoption of HomePlug specifications in home networks, little attention has been paid to providing an accurate analysis and an evaluation of~the~HomePlug~MAC~layer.

% 2. Presentation of challenges:
The vast majority of HomePlug devices employ a multiple-access protocol based on CSMA/CA that is specified by the IEEE 1901 standard~\cite{plc_st}.
This CSMA/CA mechanism resembles the CSMA/CA mechanism employed by IEEE 802.11, but with important differences in terms of complexity, performance and fairness. 
The main difference stems from the introduction of a so-called \emph{deferral counter} that triggers a redraw of the backoff counter when a station \emph{senses the medium busy}.
This additional counter significantly increases the state-space required to describe the backoff procedure. 
Moreover, as we explain in more details later, the use of the deferral counter introduces some level of coupling between the stations, which penalizes the accuracy of models based on the \emph{decoupling assumption}.
This assumption was originally proposed in the 802.11 analysis of~\cite{bianchi} and has been used in all works that have analyzed the 1901 CSMA/CA procedure so far (i.e.,~\cite{chung2006performance,vlachousigmetrics,malonehpav}). 
In this paper, we show that this decoupling assumption leads to inaccurate results, and the modeling accuracy can be substantially improved by avoiding~it.

% 3. Short-term unfairness and decoupling assumption
The decoupling assumption relies on the approximation that the backoff processes of the stations are independent and that, as a consequence, stations experience the same time-invariant collision probability, independently of their own state and of the state of the other stations~\cite{bianchi}. In addition, to analyze 1901, it has been assumed that a station senses the medium busy with the same time-invariant probability (equal to the collision probability) at any time slot~\cite{chung2006performance, vlachousigmetrics}.
In this paper, we show that the deferral counter introduces some coupling among the stations: After a station gains access to the medium, it can retain it for many consecutive transmissions before any other station can transmit. As a result, the collision and busy probabilities are not time-invariant for 1901 networks, which makes the decoupling assumption questionable.

\begin{figure}[htb!]
	\centering
	\includegraphics[scale=0.35]{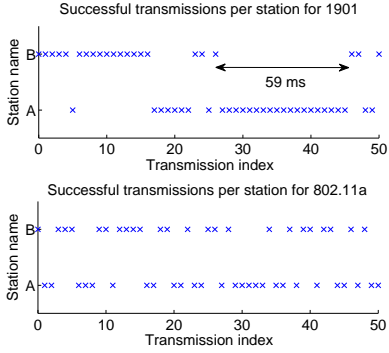}
	\caption{Testbed trace of 50 successful transmissions by two saturated stations with 1901 and 802.11a. %The experimental settings of our testbed are given in Section~\ref{sec:experiments}. 
		1901 exhibits short-term unfairness: a station holding the channel is likely to keep holding it for many consecutive transmissions, %(during several tens of ms, e.g., 59 ms as shown above), 
		which causes high dependence between the stations.
		802.11 is fairer, which makes the decoupling assumption viable in this case.
	}
	\vspace{-0.25cm}
	\label{fig:exper_fainess}
\end{figure}

\begin{figure}[t!]
	\centering
	%\vspace{-0.2cm}
	\includegraphics[scale=0.4]{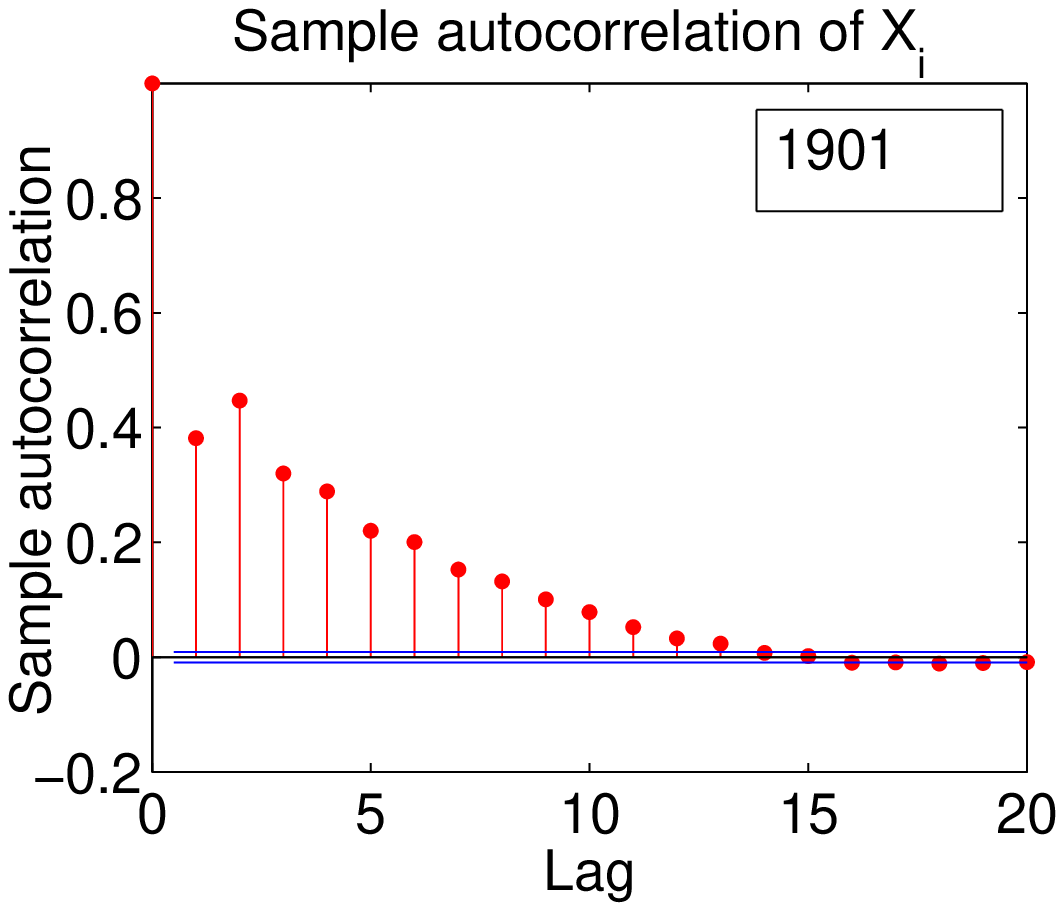}~\hspace{-2pt}\includegraphics[scale=0.4]{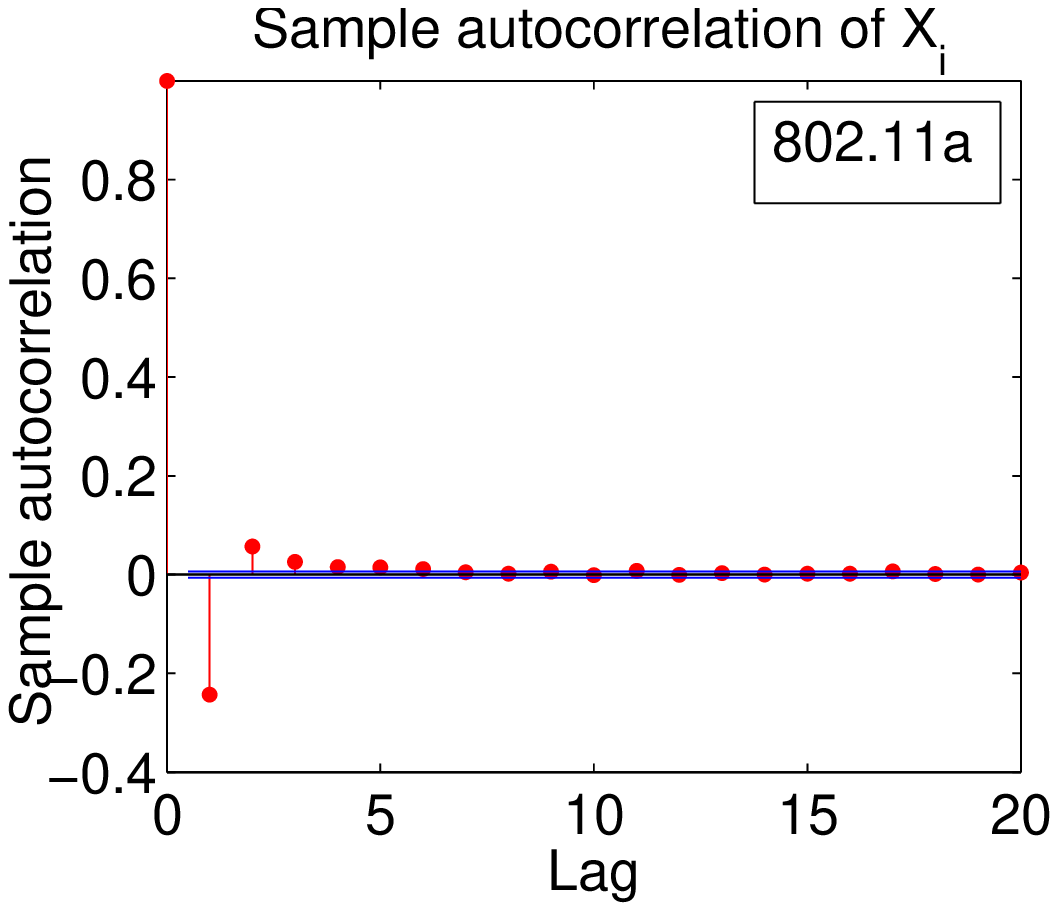}
	\caption{
		We study a testbed trace of $5\cdot10^4$ successfully transmitted packets for both 1901 and 802.11a when two saturated stations $A$ and $B$ contend for the medium. Let $X_i$ be the variable that indicates which station transmits successfully at the $i$-th transmission. We take $X_i:=1$ if $A$ transmits and $X_i:=2$ if $B$ transmits. 
		We show the autocorrelation function of $X_i,$ $1\le i \le 5\cdot10^4.$ Observe that it is positive for 1901 at lags smaller than $15$, which means that if $X_i=1$ for some $i,$ it is likely that $X_{i+1} = 1.$ For 802.11a, we have in contrast a negative value of autocorrelation at lag 1 and a positive one at lag 2, which means that if $X_i=1$ for some $i,$ it is
		very likely that $X_{i+1} = 2$ and $X_{i+2} = 1.$ 
		%This shows that 1901 is short-term unfair contrary to 802.11a.
		%
		%
		%Experimental example of the short-term unfairness of the PLC MAC with 2 saturated stations: a station is very likely hold the medium for many consecutive transmissions, while the other station waits tens of ms before transmitting again, which introduces \emph{high jitter}.
		%Short-term unfairness is related to the strong coupling between the stations that leads to inaccuracy of models that adopt the decoupling assumption for 1901.
		%We compare 1901 with 802.11 for a sample trace of 50 successful transmissions. 802.11 is short-term fair, and the decoupling is a viable assumption. 
		% The scale differences in the $x$-axis are due to different timing parameters between 802.11a and 1901.
	}
	\label{fig:exper_autocorr}
\end{figure}

Figures~\ref{fig:exper_fainess} and~\ref{fig:exper_autocorr} provide some evidence \modf{of} the coupling phenomenon described above, for a HomePlug AV testbed with two stations. While Station $A$ transmits during several consecutive slots, Station $B$ is likely to remain in a state where it has a higher probability of colliding or sensing the medium busy. $B$ is then even less likely to attempt a transmission while in this state, and it might have to wait several tens of milliseconds before the situation reverts. Thus, the collision probabilities observed by the stations are clearly time-varying, which invalidates the decoupling assumption. %Note that a consequence of this coupling is \emph{short-term unfairness}, which in turn translates into high delay variance (i.e., high jitter).

% 4. What we do in the paper:
In this paper, we propose a theoretical framework to model the CSMA/CA process of 1901 without relying on the decoupling assumption.
We first introduce a model that considers the coupling between stations and accurately captures 1901 performance.
This model is relatively compact: computing the throughput of the network only requires to solve a system of $m$ \modf{non-linear equations}, where $m$ is the number of backoff stages (the default value for 1901 is $m=4$).
We then prove that this system of equations admits a unique solution.
We confirm the accuracy of the model by using both simulations and a testbed of 7 HomePlug AV stations. 
In addition, we investigate the accuracy of our model and that of previous works relying on the decoupling assumption, showing that ours is the first model for 1901 reaching this level of accuracy.

% 5. Outline:
The remainder of the paper is organized as follows. We present the 1901 backoff process in Section~\ref{sec:background}. 
We then review the related work on MAC layer in Section~\ref{sec:related}.
We present our model for 1901 and discuss the system dynamics in Section~\ref{sec:analysis}.
We evaluate the performance of our model and discuss the decoupling assumption in Section~\ref{sec:perf_eval}.
Finally, we give concluding remarks in Section~\ref{sec:conclusion}.

\section{The CSMA/CA Protocol of IEEE 1901}
\label{sec:background}
We describe the main features of the CSMA/CA protocol used in 1901~\cite{plc_st}. We highlight in particular the mechanism that causes the strong coupling between stations and that is the main difference between 1901 and 802.11. 

The 1901 CSMA/CA procedure includes two counters: the backoff counter ($BC$) and the deferral counter ($DC$).
%In addition, there are four \emph{backoff stages}, which are determined by $BPC$.
Upon the arrival of a new packet, a transmitting station enters backoff stage $0$. 
It then draws the backoff counter $BC$ uniformly at random in $\{0,\ldots,CW_0-1\}$, where $CW_0$ denotes the contention window used at backoff stage $0$.
Similarly to 802.11, $BC$ is decreased by $1$ at each time slot if the station senses the medium to be idle (i.e., below the carrier-sensing threshold), and it is frozen when the medium is sensed busy.
In the case the medium is sensed busy, $BC$ is also decreased by $1$ once the medium is sensed idle again.
When $BC$ reaches $0$, the station attempts to transmit the packet.
Also similarly to 802.11, the station jumps to the next backoff stage if the transmission fails.
In this case, the station enters the next backoff stage.
The station then draws $BC$ uniformly at random in $\{0,\ldots,CW_i-1\}$, where $CW_i$ is the contention window used for backoff stage $i$, and repeats the process.
For 802.11, the contention window is doubled between the successive backoff stages, i.e.,~$CW_i = 2^i CW_0$. 
For 1901, $CW_i$ depends on the value of the backoff stage $i$ and the priority of the packet: There are four backoff stages, as given in Table~\ref{table:1901par}.
Also, there are two groups of priority classes (CA0/CA1 and CA2/CA3) that correspond to different values for the $CW_i$'s.

\begin{table}[!htb]
	\begin{center}
		\small
		\begin{tabular}{c  | c c | c c }
			%\cline{2-5}
			\multicolumn{1}{r}{}   &
			\multicolumn{2}{c}{Class CA0/CA1} &
			\multicolumn{2}{c}{Class CA2/CA3}\\ 
			\hline\hline
			backoff stage $i$ 
			& $CW_i$ & $d_i$ & $CW_i$ & $d_i$ \\
			% \hline \hline
			0   & 8 & 0 & 8 & 0\\
			1   & 16 & 1  & 16 & 1 \\
			2   & 32 & 3  & 16 & 3\\
			3   &64 & 15 &32 &  15\\
			\hline
		\end{tabular}
	\end{center}
	\caption{IEEE 1901 values for the contention windows $CW_i$ and the initial values $d_i$ of deferral counter $DC$, for each backoff stage $i$ and each priority class. CA0/CA1 priorities are used for best-effort traffic and
		CA2/CA3 for delay-sensitive traffic.}
	\label{table:1901par}
\end{table}

The main difference between 1901 and 802.11 is that a 1901 station might enter the next backoff stage even if it did not attempt a transmission.
This is regulated by the deferral counter $DC$, which works as follows.
When the station enters backoff stage $i$, $DC$ is set at \emph{an initial $DC$ value} $d_i$, where $d_i$ is given in Table~\ref{table:1901par} for each backoff stage $i$.
After having sensed the medium busy, a station decreases $DC$ by 1 (in addition to $BC$).
If the medium is sensed busy and $DC=0$, then the station jumps to the next backoff stage (or re-enters the last backoff stage, if it is already at this stage) and re-draws $BC$ \emph{without attempting a transmission}.
An example of such a backoff process is shown in Figure~\ref{fig:bcf}.

\begin{figure}
	\begin{center}
		\usetikzlibrary{arrows}
		\begin{tikzpicture}[>=stealth, scale=0.8]
		\tikzstyle{every node}=[font=\small]
		\draw[thick,->] (-11.2,5) -- (-11.2,-1.4);
		\node [ rotate=90] at (-11.5,1.8) {time};
		\node at (-8.4,6) {\textbf {Station $A$}};
		\node at (-3.3,6) {\textbf {Station $B$}};

		\node at (-8.5,5.3) {$CW_i$};
		\node at (-7.5,5.3) {$DC$};
		\node at (-6.5,5.3) {$BC$};
		
		\node at (-3.5,5.3) {$CW_i$};
		\node at (-2.5,5.3) {$DC$};
		\node at (-1.5,5.3) {$BC$};
		
		\draw (-9,5.5) -- (-9,-1.2);
		\draw (-8,5.5) -- (-8,-1.2);
		\draw (-7,5.5) -- (-7,-1.2);
		\draw (-6,5.5) -- (-6,-1.2);
		
		\draw (-4,5.5) -- (-4,-1.2);
		\draw (-3,5.5) -- (-3,-1.2);
		\draw (-2,5.5) -- (-2,-1.2);
		\draw (-1,5.5) -- (-1,-1.2);
		
		%%show backof stages:
		\draw [fill=green, opacity=0.3] (-11,5) rectangle (-6,-0.2);
		\draw [fill=orange, opacity=0.3] (-11,-0.2) rectangle (-6,-1.3);
		
		\draw [fill=green, opacity=0.3] (-5.8,5) rectangle (-1,3.4);
		\draw [fill=orange, opacity=0.3] (-5.8,3.4) rectangle (-1,-0.2);
		\draw [fill=green, opacity=0.3] (-5.8,-0.2) rectangle (-1,-1.3);
		
		\node at (-10.4, 5.55) {backoff};
		\node at (-10.4, 5.2) {stage $i$};
		\node at (-5.2, 5.55) {backoff};
		\node at (-5.2, 5.2) {stage $i$};
		
		\node at (-10, 2.5) {$i=0$};
		\node at (-10, -0.75) {$i=1$};
		\node at (-4.9, 4.2) {$i=0$};
		\node at (-4.9, 1.7) {$i=1$};
		\node at (-4.9, -0.75) {$i=0$};

		\node at (-8.5, 4.7) {$8$};
		\node at (-7.5, 4.7) {$0$};
		\node at (-6.5, 4.7) {$3$};
		\node at (-3.5, 4.7) {$8$};
		\node at (-2.5, 4.7) {$0$};
		\node at (-1.5, 4.7) {$5$};
		
		\node at (-8.5, 4.4) {.};
		\node at (-8.5, 4.2) {.};
		\node at (-7.5, 4.4) {.};
		\node at (-7.5, 4.2) {.};
		\node at (-6.5, 4.4) {.};
		\node at (-6.5, 4.2) {.};
		\node at (-3.5, 4.4) {.};
		\node at (-3.5, 4.2) {.};
		\node at (-2.5, 4.4) {.};
		\node at (-2.5, 4.2) {.};
		\node at (-1.5, 4.4) {.};
		\node at (-1.5, 4.2) {.};
		
		\node at (-8.5, 3.9) {$8$};
		\node at (-7.5, 3.9) {$0$};
		\node at (-6.5, 3.9) {$\mathbf{0}$};
		\node at (-3.5, 3.9) {$8$};
		\node at (-2.5, 3.9) {$0$};
		\node at (-1.5, 3.9) {$2$};
		\node [rectangle, fill=black!20] at (-7.5, 3.4) {\bf Transmission};
		
		\node at (-8.5, 2.9) {$8$};
		\node at (-7.5, 2.9) {$0$};
		\node at (-6.5, 2.9) {$7$};
		\node at (-3.5, 2.9) {$16$};
		\node at (-2.5, 2.9) {$\mathbf{1}$};
		\node at (-1.5, 2.9) {$11$};
		
		\node at (-8.5, 2.6) {.};
		\node at (-8.5, 2.4) {.};
		\node at (-7.5, 2.6) {.};
		\node at (-7.5, 2.4) {.};
		\node at (-6.5, 2.6) {.};
		\node at (-6.5, 2.4) {.};
		\node at (-3.5, 2.6) {.};
		\node at (-3.5, 2.4) {.};
		\node at (-2.5, 2.6) {.};
		\node at (-2.5, 2.4) {.};
		\node at (-1.5, 2.6) {.};
		\node at (-1.5, 2.4) {.};
		
		\node at (-8.5, 2.1) {$8$};
		\node at (-7.5, 2.1) {$0$};
		\node at (-6.5, 2.1) {$\mathbf{0}$};
		\node at (-3.5, 2.1) {$16$};
		\node at (-2.5, 2.1) {$1$};
		\node at (-1.5, 2.1) {$4$};
		\node [rectangle, fill=black!20] at (-7.5, 1.6) {\bf Transmission};
		
		\node at (-8.5, 1.1) {$8$};
		\node at (-7.5, 1.1) {$0$};
		\node at (-6.5, 1.1) {$5$};
		\node at (-3.5, 1.1) {$16$};
		\node at (-2.5, 1.1) {$\mathbf{0}$};
		\node at (-1.5, 1.1) {$3$};
		
		\node at (-8.5, 0.8) {.};
		\node at (-8.5, 0.6) {.};
		\node at (-7.5, 0.8) {.};
		\node at (-7.5, 0.6) {.};
		\node at (-6.5, 0.8) {.};
		\node at (-6.5, 0.6) {.};
		\node at (-3.5, 0.8) {.};
		\node at (-3.5, 0.6) {.};
		\node at (-2.5, 0.8) {.};
		\node at (-2.5, 0.6) {.};
		\node at (-1.5, 0.8) {.};
		\node at (-1.5, 0.6) {.};
		
		\node at (-8.5, 0.3) {$8$};
		\node at (-7.5, 0.3) {$0$};
		\node at (-6.5, 0.3) {$2$};
		\node at (-3.5, 0.3) {$16$};
		\node at (-2.5, 0.3) {$0$};
		\node at (-1.5, 0.3) {$\mathbf{0}$};
		\node [rectangle, fill=black!20] at (-2.5, -0.2) {\bf Transmission};
		
		\node at (-8.5, -0.7) {$16$};
		\node at (-7.5, -0.7) {$\mathbf{1}$};
		\node at (-6.5, -0.7) {$6$};
		\node at (-3.5, -0.7) {$8$};
		\node at (-2.5, -0.7) {$0$};
		\node at (-1.5, -0.7) {$2$};
		\node at (-8.5, -1) {.};
		\node at (-8.5, -1.2) {.};
		\node at (-7.5, -1) {.};
		\node at (-7.5, -1.2) {.};
		\node at (-6.5, -1) {.};
		\node at (-6.5, -1.2) {.};
		\node at (-3.5, -1) {.};
		\node at (-3.5, -1.2) {.};
		\node at (-2.5, -1) {.};
		\node at (-2.5, -1.2) {.};
		\node at (-1.5, -1) {.};
		\node at (-1.5, -1.2) {.};
		\end{tikzpicture}
	\end{center}
	\caption{An example of the time evolution of the 1901 backoff process with 2 saturated stations $A$ and $B$. Initially, both stations start at backoff stage $0$. Station $A$ wins the channel for two consecutive transmissions. Observe the change in $CW_i$ when a station senses the medium busy and has $DC=0.$ 
	}
	\label{fig:bcf}
\end{figure}

The deferral counter was introduced in 1901, so that 1901 can employ small contention window values -- which provide good performance for a small number of stations -- while avoiding collisions, thus maintaining good performance for a large number of stations. 
In particular, to reduce collisions, 1901 stations redraw their backoff counter when they sense a number of transmissions before their backoff counter expires; in this way, they react to a high load in the network without the need of a collision, in contrast to 802.11 that only~reacts~to~collisions.

Although the above mechanism achieves its goal, i.e., providing good performance in terms of throughput, it might lead to short-term unfairness: When a station gets hold of the channel and uses a small contention window, it is likely to transmit several frames and thus trigger the deferral counter mechanism of the other stations, which further increase their contention windows and hence reduce even more their probability of accessing the channel. Such a coupling effect penalizes the accuracy of existing models that assume that~the~backoff~processes of different stations are independent. 

\section{Related work}
\label{sec:related}
The backoff process of 802.11 can be considered as a version of 1901 where the deferral counter $DC$ never reaches $0$ (i.e., $d_i=\infty,$ for all $i$).
Hence, we first review relevant studies on 802.11 and then we present the existing work on 1901. 

\subsection{Analyses of IEEE 802.11}
Most work modeling 802.11 performance relies on the decoupling assumption, initially proposed by Bianchi in~\cite{bianchi}.
Bianchi proposes a model for single contention domains, using a discrete-time Markov chain.
Under the decoupling assumption, the collision probability experienced by all stations is time-invariant and can be found via a fixed-point equation that depends on the parameters of the protocol.
Kumar et al.~\cite{kumar} examine the backoff process of 802.11 using the same assumptions and
renewal theory. The authors also extract a fixed-point equation for the collision probability. 
The decoupling assumption has later been examined in~\cite{boudec_app},~\cite{malone} (analytically and experimentally) and found to be~valid~for~802.11.

Sharma et al.~\cite{sharma} study 802.11 without the decoupling assumption.
They analyze an $m$-dimensional chain ($m$ being the number of backoff stages) that describes the number of stations at each backoff stage.
\emph{Drift equations} capture the expected change \modf{of} the number of stations at each backoff stage between two consecutive time slots, and their equilibrium point yields the average number of stations at each backoff stage in steady state.
Similarly to~\cite{sharma}, we also use drift equations to obtain an accurate model for 1901.

\subsection{Analyses of IEEE 1901 under the Decoupling Assumption}
There are a few works analyzing the backoff mechanism of 1901 that rely on the decoupling assumption.
First, Chung et al.~\cite{chung2006performance} introduce a model using a discrete-time Markov chain similar to Bianchi's model for 802.11~\cite{bianchi}.
The additional state required to capture the effect of the deferral counter $DC$ significantly increases the complexity of the Markov chain.

Our works of~\cite{vlachousigmetrics},~\cite{vlachou2014conext} propose a simplification to the model of~\cite{chung2006performance}, reducing the Markov chain to a single fixed-point equation; by applying a similar theoretical technique to~\cite{kumar}, these papers also prove that this fixed-point equation admits a unique solution.
Being less accurate but simpler than the model introduced here, the model of~\cite{vlachou2014conext} enables us to optimize the performance of the protocol towards high throughput.

Cano and Malone~\cite{malonehpav} provide a simplification of the analysis of~\cite{chung2006performance} for computing the delay under unsaturated traffic scenarios. They also evaluate the implications of the assumption made in~\cite{chung2006performance} that the buffer occupancy probability is independent of the backoff stage at which the transmission takes place.

\section{Analysis}
\label{sec:analysis}
%Hypothesis:
In this section, we introduce our model for the 1901 CSMA/CA protocol. Our analysis relies on the following network assumptions (all of them widely used~\cite{bianchi}--\cite{vlachou2014conext}):
\begin{itemize}
	\item There is a single contention domain with $N$ stations.
	\item All stations are saturated (always have a packet to send).
	\item There is no packet loss or errors due to the physical layer, and transmission failures are only due to collisions.
	\item The stations have an infinite retry limit; that is, they never discard a packet until it is successfully transmitted.\footnote{Contrary to 802.11, 1901 does not specify a retry limit.
		However, there is a timeout on the frame transmission that is vendor specific. For instance, for the HomePlug AV devices tested in Section~\ref{sec:perf_eval}, the timeout for CA1 priority frames is 2.5 s, which is very large compared to the maximum frame duration (2.5 ms~\cite{plc_st}). Therefore, the infinite retry limit assumption is \modf{reasonable}.}
\end{itemize}

The 1901 standard introduces four different priority classes (see Section~\ref{sec:background}) and specifies that only the stations belonging to the highest contending priority class run the backoff process.\footnote{In practice, the contending priority class is decided during a so-called \emph{priority resolution phase}, using a simple system of busy tones.}
In our analysis, we follow this, and we consider a scenario in which all the contending stations use the same set of parameters (corresponding to the highest priority class).

\subsection{Baseline Model}
% We present a model that aims at studying the 1901 MAC without assuming that stations are independent.
We model the PLC network as a dynamical system that is described by the expected change in the number of stations at each backoff stage between any two consecutive time slots.
In the stationary regime, the expected number of stations at each backoff stage is constant hence,
we can compute performance metrics by finding the equilibrium of the dynamical system.

Let us now introduce the variables of our model.
%x_i, bc_i, etc:
Let $m$ be the number of backoff stages and let $n_i, 0 \leq i \leq m-1$ denote the number of stations at backoff stage $i$.
Note that $\sum_{i=0}^{m-1}n_i = N$ \modf{and $n_i\in \mathbb{N}$.}
Let us further denote with $\tau_i$ the transmission probability at stage $i$, i.e., $\tau_i$ is the probability that a station at backoff stage $i$ transmits at any given time slot.
In addition, for a given station at backoff stage $i$, we denote with $p_i$ the probability that at least one other station transmits.
We also denote with $p_e$ the probability that no station transmits (or equivalently, that the medium is idle).
Under the assumption of independence of the transmission attempts, we have $p_e = \prod_{k = 0}^{m-1}{(1-\tau_k)^{n_k}}$, therefore\footnote{The collision probability $p_i$ is defined only if at least one station is at state $i$ and is set to 0 if $n_i = 0.$}
\begin{equation}
p_i = 1-\frac{p_e}{1-\tau_i} = 1- \frac{1}{1-\tau_i}\prod_{k = 0}^{m-1}{(1-\tau_k)^{n_k}}.
\label{eq-pi}
\end{equation}

{\color{black}\emph{Model of a station}: We now model the behavior of a given station at backoff stage $i$. %We assume that the station senses a slot busy with a constant probability $p_i$. 
	We assume that the event that some other station transmits in a slot occurs with a constant probability $p_i$, independent of the station's backoff and deferral counters values.\footnote{\color{black}With this assumption, we are neglecting the coupling between the deferral counter decrements of different stations. Note, however, that this does not couple the actual transmissions, as these follow a separate random process; as a result, the coupling due to the deferral counter is somehow diluted.} Hence, this corresponds to the probability that a transmission of the given station collides, as well as to the probability that the station senses a slot busy  when it does not transmit. The rest of the backoff process of the station is modeled accurately as a function of $p_i$, drawing the station's backoff counter from a uniform distribution. With this model, we derive the probability that a station transmits and that it moves to the stage $i+1$ due to the deferral counter. These two probabilities are used in our network model presented in the next section.
}

In 1901, a station with $DC$ originally equal to $d_i$ can change its backoff stage either (i) after attempting a transmission or (ii) due to sensing the medium busy $d_i+1$ times.\footnote{A major difference between 1901 and 802.11 is that, contrary to 1901, a station using 802.11 can only adapt its backoff because of (i), not of (ii).} 
To compute the probabilities of events (i) and (ii), we introduce $x^i_k$ as the probability that a station at backoff stage $i$ jumps to backoff stage $i+1$ in $k$ or fewer time slots due to (ii).
Note that we can compute $x^i_k$ directly from $p_i$.
Let $T$ be the random variable describing the number of slots among $k$ slots during which the medium is sensed busy. Because a station at backoff stage $i$ senses the medium busy with probability $p_i$ at each time slot, $T$ follows
the binomial distribution $\text{Bin}(k,p_i).$
This yields
\begin{equation}
x^i_k = \mathbb{P}(T > d_i ) = \sum_{j = d_i+1}^{k} \binom{k}{j} p_i^{j} (1-p_i)^{k-j}.
\label{eq:def_xk}
\end{equation}

Let us denote by $bc_i$ the expected number of time slots spent by a station at backoff stage $i$. 
Now, recall that when entering stage $i$, the stations draw a backoff counter $BC$ uniformly at random in $\{0, \ldots, CW_i-1\}.$
Let $k$ denote the value of $BC.$ 
Depending on $k$, one of the following happens:
\begin{itemize}
	\item
	If $k > d_i$, then event (i) occurs with probability $(1-x^i_k),$ in which case the station spends $(k+1)$ slots in stage $i$ (the $(k+1)th$ slot being used for transmission).
	This event is illustrated by the two transmissions of Station $A$ in Figure~\ref{fig:bcf}.
	Now, (ii) occurs with probability $x^i_k$. 
	More precisely, (ii) occurs at slot $j$, for $d_i+1\le j \le k,$ with probability $(x^i_j - x^i_{j-1}),$\footnote{Observe that $(x^i_j - x^i_{j-1})$ is the difference of two complementary CDFs and denotes the probability that (ii) happens exactly at slot $j$. We have $x^i_j - x^i_{j-1} = \binom{j-1}{d_i} p_i^{d_i+1} (1-p_i)^{j-d_i-1},$ because the station senses the medium busy for $d_i$ times in any of the $j-1$ slots, and for the $(d_i+1)th$ time at the $jth$ slot.} in which case the station spends $j$ slots in stage $i.$
	\item
	If $k \leq d_i$, then (ii) cannot happen. Event (i) takes place with probability 1, which yields that the backoff counter expires 
	and that the station spends $(k+1)$ slots in stage~$i$.
\end{itemize}
By grouping all the possible cases described above, we have
\begin{align}
bc_i = &\frac{1}{CW_i}\sum_{k = d_i+1}^{CW_i-1}{\left[ (k+1)(1-x^i_{k})+\sum_{j=d_i+1}^{k}{j(x^i_j-x^i_{j-1})} \right]}\nonumber\\
&+\frac{(d_i+1)(d_i+2)}{2 CW_i}.
\label{eq-bc_i}
\end{align}
In the expression above, the first term in the outer sum captures the cases where the backoff counter expires and the station transmits at the $(k+1)$th slot (if the chosen backoff counter was $k$).
The inner sum captures all the possible cases where the backoff counter does not expire, because the station senses the medium busy $d_i+1$ times before the expiration of the backoff counter.
Finally, the last term arises because if a station chooses a backoff counter less or equal to $d_i$, then the backoff counter always expires.

Now, the transmission probability $\tau_i$ can be expressed as a function of $x^i_k$ and $bc_i,$ using the renewal-reward theorem, with the number of backoff slots spent in stage $i$ being the renewal sequence and the number of transmission attempts (i.e., 0 or 1) being the reward. Let $t_i$ be expected number of transmission attempts at stage $i$
Using a similar reasoning as for  $bc_i$,\footnote{To compute $t_i,$ observe that when Event (i) takes place we have 1 transmission attempt, and 0 attempts otherwise (Event (ii)).} $t_i$ can be computed as
\begin{equation}
t_i = \sum_{k = d_i+1}^{CW_i-1}{\frac{1}{CW_i}(1-x^i_{k})}+\frac{d_i+1}{CW_i}
\label{eq:tn}.
\end{equation}
By dividing the expected number of transmission attempts at stage $i$ with the expected time slots spent at stage $i$, $\tau_i$ is given by 
\begin{equation}
\tau_i = \frac{\sum_{k = d_i+1}^{CW_i-1}{\frac{1}{CW_i}(1-x^i_{k})}+\frac{d_i+1}{CW_i}}{bc_i}.
\label{eq-tau_i}
\end{equation}

Similarly, we define $\beta_i$ as the probability that, at any given slot, a station at stage $i$ moves to the next backoff stage because it has sensed the medium busy $d_i+1$ times (event (ii)).
$\beta_i$ is given by
\begin{equation}
\beta_i = \frac{\sum_{k = d_i+1}^{CW_i-1}{\frac{1}{CW_i}\sum_{j=d_i+1}^{k}{(x^i_j-x^i_{j-1})}}}{bc_i}.
\label{eq-beta_i}
\end{equation}
It will be very important in the following to remember that $\tau_i$ and $\beta_i$ are functions of $p_i$ (through $x^i_k$ and $bc_i$).  To simplify the exposition and the analysis of $\tau_i$ with respect to $p_i,$ we finally introduce the variable $B_i$; it is defined as $B_i \doteq 1/\tau_i - 1.$ After some computations in~\eqref{eq-tau_i}, we have 
\begin{align} 
B_{i}& = \frac{\frac{CW_{i}(CW_{i}-1)}{2}-\sum_{k = d_{i}+1}^{CW_{i}-1}{(CW_i-1-k){x^i_k}}}{CW_{i} - \sum_{k=d_{i}+1}^{CW_{i}-1}{x^i_k}}.
\label{eq-Bi-simple}
\end{align}

Our notation \modf{is} summarized in Table~\ref{table:notations}. We next study the evolution of the expected change in the number of stations at each backoff stage $i.$
\begin{table}[!htb]
	\begin{center}
		\footnotesize
		\begin{tabular}{p{0.35cm}p{0.05cm} p{7.2cm}}
			Notation & & Definition (at backoff stage $i$, $0\le i\le m-1$)\\
			\hline
			\hline
			$n_i$ & & Number of stations\\
			$p_i$ &  & Probability that at least one other station transmits at any slot \\
			$p_e$ &  & Probability that the medium is idle at any slot~(independent~of~$i$)\\
			$x_k^i$ &   & Probability that a station leaves stage $i$ due to sensing the medium busy $d_i+1$ times during $k$ slots\\
			$bc_i$ &  & Expected number of backoff slots\\
			$t_i$ & & Expected number of transmission attempts\\
			$\tau_i$ & & Probability that a station transmits at any slot\\
			$\beta_i$ && Probability that, at any slot, a station leaves stage $i$ due to sensing the medium busy $d_i+1$ times\\
			$B_i$ && $1/\tau_i -1$\\
			$F_i$ & & Expected change in $n_i$ between two consecutive slots\\
			$\bar{n}_i$ & & Expected number of stations\\
			\hline
		\end{tabular}
	\end{center}
	\caption{Notation list relevant to a station at backoff stage $i$}
	\label{table:notations}
\end{table}

\subsection{Transient Analysis of the System}
\label{sec:tr_analys}
Building on the analysis above, we now introduce our model. A key feature is that we do not assume that the stations are decoupled, as the collision probability is allowed to depend on the station's state.
To study the system, we use a vector that includes the number of stations at each backoff stage.
In particular, let $\mathbf{X}(t)=(X_0(t), X_1(t),\ldots,X_{m-1}(t))$ represent the number of stations at each backoff stage ($0, 1, \ldots, m-1$) at time slot~$t.$
We use the notation $\mathbf{n}(t)=(n_0(t), n_1(t),\ldots,n_{m-1}(t))$ to denote a realization of $\mathbf{X}(t)$ at some time slot $t$.

{\color{black}
	\emph{Network model (NM)}: To model the network, we rely on the simplifying assumption that a station transmits, or moves to the next backoff stage upon expiring the deferral counter, with a constant probability (independently of previous time slots). This is necessary as otherwise, we would need to keep track of the backoff and deferral counter values of each station and the model would become intractable. In particular, our assumptions are as follows: ($i$) a station at backoff stage $i$ attempts a transmission in each time slot with a constant probability $\tau_i(p_i)$; and ($ii$) a station at backoff stage $i$ moves to backoff stage $i+1$ due to the deferral counter expiration with a constant probability $\beta_i(p_i)$ in each time slot where it does not transmit. Both $\tau_i$ and $\beta_i$ depend on the probability $p_i$ that the station senses a slot busy, which is computed from the transmission probabilities of the other stations following~\eqref{eq-pi}.

	With the above assumptions, %\footnote{\modf{The events related to these assumptions, i.e., (i) and (ii), are disjoint.}} 
	$\mathbf{X}(t)$ is a Markov chain. The transition probabilities $\tau_i$ and $\beta_i$ depend on the state vector $\mathbf{n}(t)$ and they can be computed from~\eqref{eq-pi},~\eqref{eq-tau_i} and~\eqref{eq-beta_i}}; 
hereafter, to simplify notation, we drop the input variable $t$ from $p_i(t),$ $\tau_i(t),$ $\beta_i(t),$ and $\mathbf{n}(t)$ as the equations are expressed for any~slot~$t$.

Let now $\mathbf{F}(\mathbf{n})=\mathbb{E}[\mathbf{X}(t+1)-\mathbf{X}(t)|\mathbf{X}(t) = \mathbf{n}]$ be the expected change in $\mathbf{X}(t)$ over one time slot, given that the system is at state $\mathbf{n}$.
Function $\mathbf{F}(\cdot)$ is called the \emph{drift} of the system, and is given by
\begin{align}
\tag{DRIFT}\label{eq-drift}
&F_i(\mathbf{n}) = \\
&\quad\,\begin{cases}
{\color{black}\sum_{k = 1}^{m-1}{n_{k} \tau_{k} (1-p_{k})} - n_0 \tau_0 p_0 - n_0 \beta_0},\quad i=0\\
n_{i-1}\left( \tau_{i-1} p_{i-1} + \beta_{i-1}\right)  - n_i (\tau_i +\beta_i), \quad  0 < i < m-1\\
n_{m-2}\left( \tau_{m-2} p_{m-2} + \beta_{m-2}\right)- n_{m-1} \tau_{m-1} (1-p_{m-1}),\\
\quad\quad\quad\quad\quad\quad\quad\quad\quad\quad\quad\quad\quad\quad\quad\quad\quad\quad\text{ } i=m-1.\nonumber
\end{cases}
\end{align}
\eqref{eq-drift} is obtained by balancing, for every backoff stage, the average {\color{black}number} of stations that enter and leave this backoff stage.
In particular, $n_0$ increases by 1 only when some station transmits successfully. 
Since such a station could be in any of the other backoff stages and there are $n_k$ stations in stage $k,$ this occurs with probability
$\sum_{k=1}^{m-1}n_{k}\tau_k (1-p_{k}).$  
Similarly, $n_0$ decreases when some stations at stage $0$ are either involved in a collision (which occurs with probability $n_0 \tau_0 p_0$), 
or do not transmit and sense the medium busy $d_0+1$ times (which occurs with probability $n_0\beta_0$).
The decrease of $n_0$ in both cases is $1,$ thus the expected decrease is
equal to the sum of the two probabilities. The resulting drift $F_0$ is computed by adding all these (positive and negative) expected changes in~$n_0$.

Similarly, $F_i,\; 0 < i < m-1$ is computed 
by observing that in these backoff stages, $n_i$ decreases if and only if some stations at stage $i$ sense the medium busy or transmit.
$n_i$ increases if and only if some stations at stage $i-1$ sense the medium busy or transmit and collide. 
Finally, $n_{m-1}$ increases after some stations at stage $m-2$ experience a collision or sense the medium busy $d_{m-2}+1$ times.
It decreases only after a successful transmission at stage $m-1.$

The evolution of the expected number of stations $\bar{\mathbf{n}}(t) \doteq\mathbb{E}[\mathbf{X}(t)] $ is described by the $m$-dimensional dynamical system
\begin{equation}
%  \tag{DYNSYS}
\bar{\mathbf{n}}(t+1) = \bar{\mathbf{n}}(t) + \mathbf{F}(\bar{\mathbf{n}}(t)),
\label{eq-dynsys}
\end{equation}
where $\mathbf{F}(\bar{\mathbf{n}}(t))$ is given by~\eqref{eq-drift}.

Our model relies on the key insight that the stochastic system $\mathbf{X}(t)$ stays close to the typical state given by the equilibrium of~\eqref{eq-dynsys}, and accurate estimates of various metrics such as throughput can be obtained by assuming that the system is in this typical state at all times. However, in reality the stochastic system might stay in other states with a certain probability. {\color{black}In the following, we evaluate the accuracy of approximating our random system by a deterministic model given by \eqref{eq-dynsys}.
	
	\subsection{Accuracy of the Deterministic Model Approximation}
	\label{sec:ode}
	
	It is intuitive that the approximation becomes more accurate as the number of stations in the system grows: If the number of stations at each backoff stage is very large, the behavior of the stochastic system is expected to be close to the deterministic one given by~\eqref{eq-dynsys} due to the law of large numbers. This has been proven for 802.11 in~\cite{boudec_app,sharma,mcdonald}: By analyzing a properly scaled version of the stochastic system, these papers show that the 802.11 stochastic system converges to the deterministic model as the number of stations grows to $\infty$. In the following, we show the same result for 1901.

	As in all the previous analyses of 802.11~\cite{boudec_app,sharma,mcdonald}, we consider a scaled version of our system, $\mathbf{Y^N}(\mathcal{T})$, %$\mathbf{X}(\lfloor N t \rfloor ) / N$, 
	where time is accelerated by a factor of $N$ while the transition probabilities are scaled down by the same factor, i.e., $\mathbf{Y^N}(t/N) = \mathbf{X}(t)/N$ (this factor $N$ being equal to the number of stations):
	\begin{itemize}
		\item By scaling time, the evolution of time slots is accelerated by $N$, such that a variable at time $t$ before this operation is translated into the scaled one at time 
		$\mathcal{T} = t/N$.
		\item By scaling the transition probabilities, the evolution of each node is slowed down by a $N$.
	\end{itemize}
	
	With this scaling, the expected change of the state of the system between two consecutive time-slots is order of $1/N$, which tends to zero as $N \to \infty$. By accelerating the evolution of time-slots by $N$, the change of the system over time remains in the same order of magnitude as the original system.
	
	Following this reasoning, to scale down the transition probabilities we let the probability that a station attempts a transmission at backoff stage $i$ be $\tau_i(p_i)/N$, and the probability that it jumps to the next backoff stage due to the expiration of the deferral counter be $\beta_i(p_i)/N$. We further set $y_i(\mathcal{T})$ equal to the fraction of stations at backoff stage $i$, i.e., $y_i(\mathcal{T}) = n_i(\mathcal{T})/N$. By substituting in \eqref{eq-drift} the transition probabilities by the scaled ones, $t$ by $N \mathcal{T}$ and $n_i(\mathcal{T})$ by $N y_i(\mathcal{T})$, we obtain the following deterministic (asymptotic) system as $N \to \infty$:
	\begin{align}
	\tag{ODE}\label{ode}
	&\frac{d y_i}{d \mathcal{T}}= \\
	&\quad\,\begin{cases}
	\sum_{k = 1}^{m-1}{y_k \tau_k(\rho) (1-\rho)} - y_0 (\tau_0(\rho)\rho + \beta_0(\rho)),\quad i=0\\
	\ y_{i-1} (\tau_{i-1}(\rho) \rho + \beta_{i-1}(\rho)) - y_i (\tau_{i}(\rho) + \beta_{i}(\rho)), \\
	\quad\quad\quad\quad\quad\quad\quad\quad\quad\quad\quad\quad\quad\quad\quad\quad\text{ }  0 < i < m-1\\
	\ y_{m-2} (\tau_{m-2}(\rho)\rho + \beta_{i-1}(\rho)) - y_{m-1} \tau_{m-1}(\rho)(1-\rho),\\
	\quad\quad\quad\quad\quad\quad\quad\quad\quad\quad\quad\quad\quad\quad\quad\quad\quad\quad\text{ } i=m-1,\nonumber
	\end{cases}
	\end{align}
	where the collision probability of a station at stage $i$ is given by $\rho = %\underset{N \to \infty}{\lim}
	\lim_{N \to \infty}{1-\frac{\prod_{k=0}^{m-1}{\left(1-\tau_k(\rho)/N\right)^{n_k}}}{1-\tau_i(\rho)/N}} =  \lim_{N \to \infty}{1-\frac{\prod_{k=0}^{m-1}{\left(1-\tau_k(\rho)/N\right)^{y_k N}}}{1-\tau_i(\rho)/N}}  = 1 -{e^{- \sum_{k=0}^{m-1} y_k \tau_k(\rho)}}$. \footnote{Here, we have used the limit $\lim_{N \to \infty} \left(1-\frac{a}{N}\right)^{bN} = e^{-ab},$ for any $a,b \in \mathbb{R}^2$.} With this result, the collision probability in the asymptotic system is independent of the backoff stage $i.$
	
	The following theorem shows that the stochastic system under study converges to the deterministic model given above as $N \to \infty$, which confirms that the proposed analysis becomes very accurate as the number of stations grows large.
	
	\begin{thm} As $N \to \infty$, the scaled random system $\mathbf{Y^N}(\mathcal{T})$ converges to the deterministic process $\mathbf{y}(\mathcal{T})$ given by \eqref{ode}.
	\end{thm}
	\begin{IEEEproof} 
		The proof follows from the work by Benaïm and Le Boudec\cite{boudec}; according to this paper, it is sufficient to verify that the system analyzed satisfies the five assumptions given in \cite{boudec}, referred to as H1--H5. As our system does not have the so-called \mquote{common resource} (an additional entity with which the stations can interact), assumptions H1 and H4 are not applicable here. By taking the \mquote{vanishing intensity} $\epsilon(N) = 1/N$, it follows that $
		\lim_{N \to \infty} \mathbf{F}(\mathbf{y})/\epsilon(N)$ can be expressed as a function of $\tau_i(\rho)$ and $\beta_i(\rho)$\footnote{Observe that $ \mathbf{F}(\mathbf{y})/\epsilon(N)$ involves products of the form $(n_i \cdot \tau_i / N)\cdot N,$ because of the scaled transition probabilities, and all the $N$ terms cancel out.} hence, assumption H2 (\mquote{\emph{Intensity vanishes at a rate $\epsilon(N)$}}) is satisfied.
		
		To validate assumption H3 (\mquote{\emph{Second moment of number of object transitions per time slot has an upper bound of order $N^2\epsilon^2(N) $}}), we proceed as follows. Let $\alpha = \max_{i,p_i}(\tau_{i}(1-p_i),\tau_i p_i+\beta_i)$. % (over all backoff stages and feasible values of $p_i$). 
		Note that $\alpha < 1$\footnote{We have $\alpha < 1$, given that $\tau_{i}(1-p_i) < 1$ and $\tau_i p_i +\beta_i < 1$ (the latter follows from $\tau_i +\beta_i = 1/bc_i < 1$).}. %The probability that $j$ stations change their state is smaller than $(\alpha/N)^j$. Thus,
		Then, the probability that a station changes its state is upper bounded by $\alpha/N$, and the number of stations that change their state is stochastically upper bounded by a random variable $W_N$ that follows the binomial distribution $\text{Bin}(N,\alpha/N)$\footnote{Note that, under our network model, a station changes its state independently of the transitions of the other stations.}. Thus, we have
		$
		\mathbb{E}[W_N^2] = \mathbb{E}[W_N]^2 + \operatorname{Var}[W_N] < \alpha^2 + \alpha,
		$ 
		which implies that assumption H3 is satisfied.
		
		Finally, both $\tau_i$ and $\beta_i$ are smooth functions of $p_i$ that in turn is a smooth function of the $n_i$'s, because all the aforementioned functions are continuous and continuously differentiable. Hence, the transition probabilities are smooth functions of the $y_i$'s. In addition, the transition probabilities chosen are also smooth functions of $N$. This also holds for the boundaries of the transition probabilities, including the case when $N \to \infty$. Therefore, assumption H5 (\mquote{\emph{$\mathbf{F}(\mathbf{n})$ is a smooth function of $1/N$ and $\mathbf{n}$}}) is also satisfied.
		
	\end{IEEEproof}
}

\subsection{Steady-State Analysis of the System}
\label{sec:analysis_stat}

We next study the the system under steady state. To obtain the average number of stations at each backoff stage in steady state, we compute the equilibrium point(s) of system~\eqref{eq-dynsys} corresponding to the stationary regime. 
To compute the equilibrium point(s) of~\eqref{eq-dynsys}, we impose $\mathbf{F}(\bar{\mathbf{n}}) = \mathbf{0},$ which yields
\begin{align*}
\bar{n}_i &= \left(\frac{\tau_{i-1} p_{i-1} + \beta_{i-1}}{\tau_i + \beta_i}\right) \bar{n}_{i-1}, \quad 1\leq i \leq m-2,\\
\bar{n}_{m-1} &= \left(\frac{\tau_{m-2} p_{m-2} + \beta_{m-2}}{\tau_{m-1}(1-p_{m-1})}\right) \bar{n}_{m-2}.
\end{align*}
Let us define
\begin{align}
&K_0 \doteq 1,\, K_i \doteq \frac{\tau_{i-1} p_{i-1} + \beta_{i-1}}{\tau_i + \beta_i}, \quad 1\leq i \leq m-2,
\nonumber\\
\label{eq-ki}
&K_{m-1} \doteq \frac{\tau_{m-2} p_{m-2} + \beta_{m-2}}{\tau_{m-1}(1-p_{m-1})}.
\end{align}
Since $\sum_{i=0}^{m-1}{\bar{n}_i} = N$, the equilibrium $\hat{\mathbf{n}}$ of~\eqref{eq-dynsys} is given by 
\begin{align}\tag{EQ}
\hat{n}_i = \frac{N\prod_{j = 0}^{i}{K_j}}{\sum_{k = 0}^{m-1}{\prod_{j = 0}^{k}{K_j}}},  \, 0\leq i \leq m-1\label{eq-ni}.
%&p_i = 1- \frac{1}{1-\tau_i}\prod_{k = 0}^{m-1}{(1-\tau_k)^{n_k}}, 0\le i \le m-1.\nonumber
\end{align}

Recall that $\tau_i$ and $\beta_i$ are functions of $p_i$, given by (\ref{eq-tau_i}) and (\ref{eq-beta_i}). Thus, the $\hat{n}_i$'s in (\ref{eq-ni}) are also functions of $p_i,\,0 \leq i \leq m-1.$ 
From the above, substituting (\ref{eq-ni}) in~\eqref{eq-pi} yields a system of $m$ equations with $m$ unknowns $p_i$ for $0 \leq i \leq m-1$.

After solving the equations for finding the steady-state number of nodes $\hat{n}_0, \ldots, \hat{n}_{m-1}$ at each backoff stage, we can compute the throughput of the network as follows. 
The probability that a slot is idle is $p_e.$ 
The probability of a successful transmission of a station at stage $i$ is $\tau_i (1-p_i)$.
Therefore, the probability $p_s$ that a slot contains a successful transmission is given by
$
p_s = \sum_{i=0}^{m-1}\hat{n}_i\tau_i (1-p_i),
$
assuming that $\mathbf{n}$ remains in a neighborhood of the equilibrium point $\hat{\mathbf{n}}.$
Let $p_c$ denote the probability that a slot contains a collision. We have 
$
p_c=1-p_e-p_s.
$
We now have enough information to compute the normalized throughput $S$ of the network as
\begin{equation}
S=\frac{p_s D}{p_s T_s+ p_c T_c+p_e \sigma},
\label{eq:thr}
\end{equation}
where $D$ is the frame duration, $T_s$ is the duration of a successful transmission, $T_c$ is the duration of a collision, and $\sigma$ is the time slot duration.
In Section~\ref{sec:perf_eval}, we evaluate the stationary regime of the system, and show that our model is very accurate for a wide range of configurations.

\subsection{Uniqueness of the Equilibrium Point}
\label{sec:uniqueness}
{In this subsection we prove that, as long as the configuration of $CW_i$'s and $d_i$'s is chosen such that the sequence $\tau_i$ is decreasing with $i$ for any $n_i$ distribution, then the equilibrium point given by~\eqref{eq-ni} is unique. 
	We argue that such a condition should be met by any sensible configuration of $CW_i$'s and $d_i$'s.} 
The argument is as follows. Jumping to the next backoff stage is an indication of high contention, either because of a collision or a sequence of busy slots. Therefore, in this case $\tau_i$ should decrease with $i$, and the high contention should be dissolved by reducing the aggressiveness of the sources. 
Note that similar studies for the 802.11 MAC protocol~\cite{boudec_app,kumar} require the same sufficient condition (i.e., $\tau_i$ decreasing with $i$) for the model to admit a unique solution. To simplify the exposition, we define this condition as follows.
\begin{equation}\tag{COND}
\tau_i > \tau_{i+1},\, 0 \le i \le m-2.
\label{eq-cond}
\end{equation}
We now prove, in Theorem~\ref{thm:unicity}, that if~\eqref{eq-cond} is satisfied, the equilibrium point given by~\eqref{eq-ni} is unique. 
\begin{thm}
	The system of equations formed by (\ref{eq-ni}) and (\ref{eq-pi}) for $0 \le i \le m-1$ has a unique solution if (\ref{eq-cond}) is satisfied.
	\label{thm:unicity}
\end{thm}
\begin{IEEEproof}
	Recall that $p_e = \prod_{k = 0}^{m-1}{(1-\tau_k)^{\hat{n}_k}}$. For any value of $p_e,$ $\tau_i$ can be computed from the fixed-point equation that results from combining (\ref{eq-pi}) (i.e., $p_i = 1-p_e/(1-\tau_i)$) with (\ref{eq-tau_i}), where (\ref{eq-tau_i}) is expressed as a function of $p_i$ through~\eqref{eq:def_xk}. Hence, $\tau_i$ can be computed as a function of $p_e$, and so can $p_i,$ and $\beta_i.$ Now, $\hat{n}_i$ can also be computed as a function of $p_e$ using~\eqref{eq-ni}. Let 
	$\Phi(p_e) \doteq \prod_{k = 0}^{m-1}{(1-\tau_k(p_e))^{\hat{n}_k(p_e)}}.$ Then, a solution of (\ref{eq-ni}) has to satisfy the following equation:
	\begin{equation}
	p_e = \Phi(p_e).
	\label{eq-pe-unique}
	\end{equation}

	It can be seen that~(\ref{eq-pe-unique}) has at least one fixed-point. {$\Phi(p_e)$ is defined in $[0,1-\tau_{max}],$ where $\tau_{max} := 2/(CW_0+1)$ is the maximum transmission probability at stage 0. Observe that $\Phi(0) > 0$ and $\Phi(1-\tau_{max})< 1-\tau_{max}$ thus, by the intermediate value theorem, $\Phi(p_e)$ has at least one fixed-point in $[0,1-\tau_{max}]$.
		We now show that (\ref{eq-pe-unique}) has only one fixed-point.} 
	To this end, we show that $\Phi(p_e)$ is monotonically decreasing with $p_e$.
	The derivative of $\Phi(p_e)$ can be written as
	\begin{equation}{
		\frac{d \Phi(p_e)}{d p_e} = \sum_{j = 0}^{m-1} {\left(\frac{\partial \Phi}{\partial p_j}\frac{d p_j}{d p_e}+\frac{\partial \Phi}{\partial \beta_j}\frac{d \beta_j}{d p_e}+\frac{\partial \Phi}{\partial \tau_j}\frac{d \tau_j}{d p_e}\right)}.}
	\label{eq-derphi}
	\end{equation}
	We now examine separately each of the partial derivative products of~\eqref{eq-derphi} with respect to $p_j,\,\beta_j$ and $\tau_j.$
	To prove the theorem, we rely on our analysis in the Appendix.
	First, Lemmas~\ref{lemma:tau_bci} and \ref{lemma:taui_vs_pe} imply respectively that {$d p_j /d \tau_j <0$ and $ d \tau_j/ d p_e <0.$}   Because
	${
		d p_j/d p_e  = (d p_j /d \tau_j)\cdot( d \tau_j/ d p_e),}
	$
	we have ${d p_j / d p_e <0.}$ Also, from Lemma~\ref{lemma:pi_inc}, we have $\partial \Phi/\partial p_j > 0.$ Thus, the first product of partial derivatives in~\eqref{eq-derphi} is negative for all $j$.
	Second, from Lemma~\ref{lemma:betai_inc}, we have $\partial \Phi/\partial \beta_j \ge 0.$ Now, Corollary~\ref{cor:betai} states that {$d \beta_j/d p_j>0$} and we have shown above that {$d p_j / d p_e <0.$} Hence, we have {$d \beta_j/d p_e < 0$}
	thus, the second product of partial derivatives in~\eqref{eq-derphi} is also negative.
	Third, from Lemma~\ref{lemma:taui_inc} we have $\partial \Phi/\partial \tau_j < 0,$ and from Lemma~\ref{lemma:taui_vs_pe} we have {$d \tau_j / d p_e > 0$.}  
	We have shown that all the partial derivative products of~\eqref{eq-derphi} are negative, so $\Phi(p_e)$ is monotonically decreasing with $p_e.$
	
	Since~\eqref{eq-derphi} is strictly negative and~\eqref{eq-pe-unique} admits at least one fixed-point, there exists a unique value~for $p_e$ that solves~\eqref{eq-pe-unique}. Computing the corresponding value for~$p_i$ by~\eqref{eq-pi}, we have a solution to~\eqref{eq-ni}.
	The uniqueness of the solution then follows from the fact that all relationships between $\tau_i,$ $\beta_i,$ $p_i$ and $p_e$ are bijective (because all $\tau_i,$ $\beta_i,$ and $p_e$ are monotone functions of $p_i$, as shown in the Appendix), and any solution must satisfy~\eqref{eq-pe-unique}, which (as we have shown) has only one solution.
	~\end{IEEEproof}
We next provide some configuration guidelines for $(CW_i,d_i)$ that ensure that~\eqref{eq-cond} is satisfied. In Section~\ref{sec:perf_eval}, we discuss a counterexample of a configuration that does not satisfy~\eqref{eq-cond} and does not yield a unique solution to the system of equations formed by (\ref{eq-ni}) and (\ref{eq-pi}). 
\subsection{Protocol Configurations Satisfying~\eqref{eq-cond}}
Before showing in Theorem~\ref{thm:taui_dec} that~\eqref{eq-cond} is satisfied for a wide range of configurations, we prove a useful lemma. Note that compared to 802.11, where $\tau_i$ is a function of only $CW_i,$ the analysis here is substantially more challenging, because $\tau_i$ is a function of  $CW_i,\,d_i,$ and $p_i.$

We have to investigate the relationship between $\tau_i$ and $\tau_{i+1}.$ Recall that these two transmission probabilities are functions of two different collision probabilities $p_i$ and $p_{i+1},$ respectively, which makes the analysis challenging.
Assume that the collision probability is the same for two successive backoff stages $i$ and $i+1$, and is equal to $p_i$.
Under this hypothesis, in Lemma~\ref{lm:taui_dec}, we show that if $\tau_i(p_i) > \tau_{i+1} (p_i)$, $\forall p_i\in [0,1],$ then $\tau_i(p_i) > \tau_{i+1}(p_{i+1}),$ for any pair $p_i, p_{i+1}$ that satisfies \eqref{eq-pi}.
In Theorem~\ref{thm:taui_dec}, we provide some sufficient conditions to guarantee that $\tau_i(p_i) > \tau_{i+1} (p_i)$ is satisfied for all $ p_i\in [0,1]$ and $0 \le i < m-1$; from Lemma~\ref{lm:taui_dec}, this implies that~\eqref{eq-cond} is satisfied.

\begin{lm}
	\ Let $p_i^s $ be a value of the collision probability at stage $i.$ Then, if $\tau_i (p_i^s) > \tau_{i+1} (p_{i}^s)$ for all $p_i^s \in [0,1],$ we have $\tau_i (p_i) > \tau_{i+1} (p_{i+1})$
	for any $n_i$ distribution. %$p_i$ and $p_{i+1}$ satisfying \eqref{eq-pi}.
	\label{lm:taui_dec}
\end{lm}
\begin{IEEEproof}\
	The proof goes by contradiction. Let us assume that there exists a {solution $\mathbf{n^s}$, such that the corresponding values of $\tau_i^s$, $p_i^s$, $\tau_{i+1}^s$ and $p_{i+1}^s,p_e^s$ satisfy} $\tau_i^s (p_i^s) < \tau_{i+1}^s (p_{i+1}^s) $ and (consequently) $p_i^s > p_{i+1}^s$, which contradicts what we would like to prove. Note that, for any $n_i$ distribution, $p_i$ and $p_{i+1}$ satisfy \eqref{eq-pi}. Due to~\eqref{eq-pi}, we have 
	\begin{equation}
	\frac{1-p_i^s}{1-p_{i+1}^s} = \frac{1-\tau_{i+1}^s}{1-\tau_{i}^s}.
	\label{eq-sol}
	\end{equation}
	Let us fix $\tau_i$ and $p_i$ to the values given by the solution described above, and vary $p_{i+1}$ by choosing different values of $\gamma$, defined as  {$1-p_{i+1} \doteq \gamma (1-p_i^s)$}. For each $\gamma$, we first compute $\tau_{i+1}$ that corresponds to this $p_{i+1}$, and then we compute the expression $(1-\tau_i^s)/(1-\tau_{i+1})$ that results from this $\tau_{i+1}$ and the (fixed) $\tau_{i}^s$ value. Then, if such a solution $s$ exists, there must be some value of $\gamma \geq 1$ for which 
	\begin{equation}
	f(\gamma) \doteq \frac{1-\tau_i^s}{1-\tau_{i+1}(\gamma)} = \gamma, \nonumber
	\label{eq-fsol}
	\end{equation}
	because of~\eqref{eq-sol} and the definition of $\gamma$. Next, we show that such a $\gamma$ does not exist, which contradicts our initial assumption.

	By hypothesis, for $\gamma = 1$ we have $p_{i+1} = p_i^s,$ so $\tau_i(p_i^s) > \tau_{i+1}(p_i^s),$ and $f(1) <1.$ A sufficient condition to ensure that there exists no $\gamma > 1$ value for which $f(\gamma) = \gamma$ is that the derivative of $f(\gamma)$, i.e.~$d f(\gamma)/d \gamma$, does not exceed 1 in the region $\gamma \geq 1$. To prove this, we proceed as follows.
	\begin{align} 
	\frac{d f(\gamma)}{d \gamma} & = \frac{1-\tau_i^s}{(1-\tau_{i+1})^2} \frac{d \tau_{i+1}}{d \gamma} =
	\frac{1-\tau_i^s}{(1-\tau_{i+1})^2}\frac{d \tau_{i+1}}{d p_{i+1}}\frac{d p_{i+1}}{d \gamma} \nonumber\\
	&=  -\frac{(1-\tau_i^s)(1-p_i^s)}{(1-\tau_{i+1})^2}\frac{d \tau_{i+1}}{d B_{i+1}}\frac{d B_{i+1}}{d p_{i+1}} \nonumber\\
	& =\frac{(1-\tau_i^s)(1-p_i^s)\tau_{i+1}^2}{(1-\tau_{i+1})^2}\frac{d B_{i+1}}{d p_{i+1}}  \nonumber\\
	&\overset{\gamma \ge1} {\leq} \frac{(1-p_{i+1})\tau_{i+1}^2}{(1-\tau_{i+1})^2}\frac{d B_{i+1}}{d p_{i+1}} = 
	\frac{(1-p_{i+1})}{B_{i+1}^2}\frac{d B_{i+1}}{d p_{i+1}}.\nonumber
	\end{align}
	
	From the above, it is sufficient to prove $d B_{i+1}/d p_{i+1} \leq B_{i+1}^2/(1-p_{i+1})$. 
	This is shown in Corollary~\ref{cor:bi-der} in the Appendix for $ p_{i+1} \in [0,1).$ For $p_{i+1} = 1$, we also have $p_{i} = 1$ by~\eqref{eq-pi}. Thus, a solution $s$ cannot exist and $\tau_i(p_i) > \tau_{i+1} (p_{i+1})$.
\end{IEEEproof}

The following theorem provides some sufficient conditions on the $(CW_i,\, d_i)$ configurations that ensure that~\eqref{eq-cond} holds. 
Notably, Lemma~\ref{lm:taui_dec} can be employed to show that~\eqref{eq-cond} holds for more configurations than the ones covered by the theorem; indeed, it is sufficient to show that the configuration satisfies the hypothesis of the Lemma~\ref{lm:taui_dec} for all $0\le i \le m-2$.

\begin{thm}
	~\eqref{eq-cond} holds if the following condition is satisfied for $0 \le i \le m-2$
	\begin{equation}\label{eq-cw-condition}
	CW_{i+1} > \begin{cases} CW_i, &\mbox{if } d_{i+1} = d_i\\
	2 CW_i - d_{i} -1, &\mbox{otherwise.}\end{cases}
	\end{equation}
	\label{thm:taui_dec}
\end{thm}
\begin{IEEEproof}
	We analyze two cases: 1) $d_{i+1} = d_i$; 2) $d_{i+1} \neq d_i.$
	
	1) We start for the case $d_{i+1} = d_i$. 
	By using Lemma~\ref{lm:taui_dec}, we need only to prove that $\tau_{i+1}(p_i) < \tau_{i} (p_i).$ If this is satisfied for $CW_{i+1} = CW_i + 1$, by using induction it is easy to see that it holds for any $CW_{i+1} > CW_i.$
	Now, as $\tau_i = 1/ (B_i+1)$, it is sufficient to show that $B_{i+1} > B_i$ when $CW_{i+1} = CW_i + 1.$ 
	
	If $B_{i+1}>B_i$ holds for $CW_{i+1} = CW_i + 1$, by using induction it is easy to see that it holds for any $CW_{i+1} > CW_i.$ Thus, we prove the corollary for $CW_{i+1} = CW_i + 1$. 
	
	Because $d_i = d_{i+1},$ $p_i = p_{i+1}$ and by using~\eqref{eq:def_xk}, we have $x^i_k = x_k^{i+1}$ for all $d_i + 1 \leq k \leq CW_i - 1$. Given this, we have 
	\begin{align*} 
	B_{i+1} - B_i &\\
	{=}\: &\frac{(CW_i - \sum_{j=d_{i}+1}^{CW_{i}-1}{x^i_j})^2}{({CW_{i}+1 - \sum_{j=d_{i}+1}^{CW_{i}}{x^i_j}})({CW_{i} - \sum_{j=d_{i}+1}^{CW_{i}-1}{x^i_j}})}\nonumber\\
	& {-}\:\frac{(1-x^i_{CW_i})\frac{CW_i(CW_i-1)}{2} }{({CW_{i}+1 - \sum_{j=d_{i}+1}^{CW_{i}}{x^i_j}})({CW_{i} - \sum_{j=d_{i}+1}^{CW_{i}-1}{x^i_j}})\nonumber}\\
	& {+}\:\frac{(1-x^i_{CW_i})\sum_{j = d_{i}+1}^{CW_{i}-1}{(CW_i-1-j){x^i_j}}}{({CW_{i}+1 - \sum_{j=d_{i}+1}^{CW_{i}}{x^i_j}})({CW_{i} - \sum_{j=d_{i}+1}^{CW_{i}-1}{x^i_j}})\nonumber}.
	\end{align*}
	By the definition of $x^i_k$ in~\eqref{eq:def_xk}, we have $x^i_{CW_i} \ge x^i_k,$ $d_i + 1 \le k \le CW_i - 1,$ with equality at $p_i = 0,\, 1.$ This yields $\sum_{j = d_i + 1} ^ {CW_i - 1} x^i_j \le (CW_i - d_i -1) x^i_{CW_i} < CW_i x^i_{CW_i}.$ We thus have
	\begin{align*} 
	B_{i+1} -  B_i   & \\
	{>}\: & \frac{CW_i (1-x^i_{CW_i}) (CW_i - \sum_{j=d_{i}+1}^{CW_{i}-1}{x^i_j})}{({CW_{i}+1 - \sum_{j=d_{i}+1}^{CW_{i}}{x^i_j}})({CW_{i} - \sum_{j=d_{i}+1}^{CW_{i}-1}{x^i_j}})}\nonumber\\
	&{-}\:\frac{(1-x^i_{CW_i})\frac{CW_i(CW_i-1)}{2}}{({CW_{i}+1 - \sum_{j=d_{i}+1}^{CW_{i}}{x^i_j}})({CW_{i} - \sum_{j=d_{i}+1}^{CW_{i}-1}{x^i_j}})\nonumber}\\
	&{+}\:\frac{(1-x^i_{CW_i})\sum_{j = d_{i}+1}^{CW_{i}-1}{(CW_i-1-j){x^i_j}}}{({CW_{i}+1 - \sum_{j=d_{i}+1}^{CW_{i}}{x^i_j}})({CW_{i} - \sum_{j=d_{i}+1}^{CW_{i}-1}{x^i_j}})\nonumber}\\
	{}={} &\frac{(1-x^i_{CW_i})(\frac{CW_i(CW_i+1)}{2} - \sum_{j=d_{i}+1}^{CW_{i}-1}{(j+1)x^i_j})}{({CW_{i}+1 - \sum_{j=d_{i}+1}^{CW_{i}}{x^i_j}})({CW_{i} - \sum_{j=d_{i}+1}^{CW_{i}-1}{x^i_j}})}\\
	{}\ge{} & 0 \nonumber,
	\end{align*}
	where the last inequality holds because $x^i_j \le 1,$ for all $i,j.$ 
	
	2) We now look at the case $d_{i+1} \neq  d_i$. The result for this case follows from Lemma~\ref{lemma:tau_bci} in the Appendix. Using this, the minimum value of $B_{i}$ is $B_{i}^{min} \doteq (CW_{i}-1)/2$ at $p_{i} = 0$, and its maximum value is $B_{i}^{max} \doteq CW_{i}-d_i/2-1$ at $p_{i} = 1$. Setting $CW_{i+1} > 2 CW_i - d_{i} -1$ yields $B_{i+1}^{min} > B_{i}^{max}.$ Hence, $B_{i+1} > B_i$ for all $p_{i} \in [0,1]$, $p_{i+1} \in [0,1]$.
	
\end{IEEEproof}

Observe that, from Table~\ref{table:1901par}, the above constraints on $CW_i$ and $d_i$ are compliant with the standard, except for the class CA2/CA3 at backoff stage $i=1$. The results obtained in this paper suggest that it might be worth to revisit the configuration of this priority class; indeed, for the proposed configuration of CA2/CA3 we have $\tau_2 > \tau_1$ and~\eqref{eq-cond} does not hold.

We next discuss whether~\eqref{eq-cond} is sufficient for the global asymptotic stability of~\eqref{eq-ni}.

\subsection{Global Asymptotic Stability and Convergence}

In addition to showing that our system has only one equilibrium point, it is also interesting to show that it converges to this equilibrium point from all possible initial states. 
In the following, we study the global stability of the system to assess its convergence to the equilibrium. We first study analytically the system \eqref{ode} given in Section~\ref{sec:ode}, and present two theorems that guarantee that it is globally asymptotically stable for $m = 2$ and $m = 3$, respectively. Then, we provide additional numerical results that show that the system given by~\eqref{eq-dynsys} converges for a wide range of values of $m$ as well as of the other system parameters.

\begin{thm} If~\eqref{eq-cond} is satisfied, the system \eqref{ode} is globally asymptotically stable for $m = 2$.
\end{thm}
\begin{IEEEproof}
	Let $\gamma(t) = \tau_0 y_0(t) + \tau_1 y_1(t).$ Then, $\rho = 1-e^{-\gamma}$ and~\eqref{ode} for $m=2$ is given by
	\begin{align*}
	\frac{d y_0(t)}{d t} & = -y_0 (\tau_0(1-e^{-\gamma})+\beta_0) + y_1 \tau_1 e^{-\gamma} \nonumber\\
	\frac{d y_1(t)}{d t} & = y_0 (\tau_0(1-e^{-\gamma})+\beta_0) - y_1 \tau_1 e^{-\gamma} \nonumber,
	\end{align*}
	where the $\tau_i$'s and $\beta_i$'s are functions of $\rho,$ or equivalently of $\gamma.$
	
	Let us consider the Lyapunov function $L(\mathbf{y}) = (y_0(t)-\hat{y}_0)^2 + (y_1(t)-\hat{y}_1)^2$, where $\hat{y}_0$ and $\hat{y}_1$ are the values of $y_0(t)$ and $y_1(t)$ at the equilibrium point. If at some time $t$ we have $y_0(t) > \hat{y}_0$, this implies $y_1(t) < \hat{y}_1$ (since $y_0(t) + y_1(t) = 1$) and $\gamma(t) > \hat{\gamma}$. The latter can be seen by contradiction. Let us assume $\gamma(t) < \hat{\gamma}$. Then, from $\gamma = -\ln(1-\rho)$ and Lemma \ref{lemma:tau_bci} we have $\tau_0 > \hat{\tau}_0$ and $\tau_1 > \hat{\tau}_1$. Thus, $\gamma = \tau_0 y_0(t) + \tau_1 y_1(t) > \hat{\tau}_0 y_0(t) + \hat{\tau}_1 y_1(t) = y_0 (\hat{\tau}_0-\hat{\tau}_1) + \hat{\tau}_1 >  \hat{y}_0 (\hat{\tau}_0-\hat{\tau}_1) + \hat{\tau}_1 = \hat{\tau}_0 \hat{y}_0(t) + \hat{\tau}_1 \hat{y}_1(t) = \hat{\gamma}$, since we have $\hat{\tau}_0 > \hat{\tau}_1 $ from~\eqref{eq-cond}. This contradicts the initial assumption.
	
	Given $\gamma(t) > \hat{\gamma}$, we have $\tau_0(1-e^{-\gamma})+\beta_0 > \hat{\tau}_0(1-e^{-\hat{\gamma}})+\hat{\beta}_0$. This can be seen as follows. By employing a similar reasoning to Lemma \ref{lemma:tau_bci}, we have $\partial (\tau_0+\beta_0)/\partial \gamma > 0$. We also have $- \partial (\tau_0 e^{-\gamma})/\partial \gamma > 0$ from Lemma~\ref{lemma:tau_bci}. Then, adding both expressions we obtain $\partial (\tau_0(1-e^{-{\gamma}})+\beta_0)/\partial \gamma > 0$. Given $\gamma(t) > \hat{\gamma}$, we also have $\tau_1 e^{-\gamma} < \hat{\tau}_1 e^{-\hat{\gamma}}$. Thus,
	\begin{align*}
	\frac{d y_0(t)}{d t} & = -y_0 (\tau_0(1-e^{-\gamma})+\beta_0) + y_1 \tau_1 e^{-\gamma} \nonumber\\
	& < - y_0 (\hat{\tau}_0(1-e^{-\hat{\gamma}}) +\hat{\beta}_0) + y_1 \hat{\tau}_1 e^{-\hat{\gamma}} \nonumber\\
	& < -\hat{y}_0 (\hat{\tau}_0(1-e^{-\hat{\gamma}})+\hat{\beta}_0) + \hat{y}_1 \hat{\tau}_1 e^{-\hat{\gamma}} = 0.
	\end{align*}
	
	Since $d y_0(t)/d t + d y_1(t)/d t = 0$, this in turn implies $d y_1(t)/d t > 0$. Putting all this together yields
	\begin{align*}
	\frac{d L(\mathbf{y})}{d t} = 2(y_0(t)-\hat{y}_0)\frac{d y_0(t)}{d t} + 2(y_1(t)-\hat{y}_1)\frac{d y_{1}(t)}{d t} < 0.
	\end{align*}
	
	Following a similar reasoning, it can be seen that if $y_0(t) < \hat{y}_0$, then $d y_0(t)/d t > 0$ and $d y_1(t)/d t < 0$. As a consequence, we have $d L(\mathbf{y})/d t < 0$ also in this case. Therefore, the system is globally asymptotically stable.
	%$lim_{t \to \infty} \mathbf{y}(t) = \mathbf{\hat{y}}$.
\end{IEEEproof}

\begin{thm} If~\eqref{eq-cond} is satisfied, the system \eqref{ode} is globally asymptotically stable for $m = 3$. \label{thm:ode3}
\end{thm}
\begin{IEEEproof}
	See the Appendix.
\end{IEEEproof}

In order to show the convergence of the system for other values of $m$, we have conducted a comprehensive numerical study for the dynamical system given by~\eqref{eq-dynsys}, comprising all the values of the parameters $CW_i$, $m$ and $d_i$ that satisfy \eqref{eq-cond} within the ranges $CW_{i} = \{8,16,32,64\}$, $m = \{3,4,5,6\}$ and $d_i = \{0,1,2,3,4,5,6,7,8,9,10,15,20,25,30\}$. For each configuration, we have randomly chosen 100 different initial points and evaluated the trajectory of the system until it converges (with an error of $10^{-8}$). In total, around $10^6$ tests of convergence have been conducted, and in every test, the system converges to the equilibrium given by \eqref{eq-ni}. 
All these (numerical and theoretical) results provide very strong evidence of the convergence of the system.

\section{Performance Evaluation}
\label{sec:perf_eval}
In this section, we evaluate the performance of 1901 under different configurations and scenarios as follows. We first validate our simulator by using testbed experiments. We then compare the accuracy of our model against a model that relies on the decoupling assumption, called ``D.A.'' model~\cite{vlachousigmetrics}. Furthermore, we evaluate other aspects of our model, such as the performance of the configurations that do not satisfy~\eqref{eq-cond} and the accuracy of our model in the transient regime.

We conduct simulations for a wide range of configurations $(CW_i,\, d_i),$ {comprising the parameters recommended by the 1901 standard as well as more broad configurations, following the recommendations of~\cite{vlachou2014icnp} and~\cite{vlachou2014conext}.} We consider the following timing parameters. We use the same time slot duration and timing parameters as specified in the standard (see Table~\ref{table:1901parslots}). 
The PLC frame transmission has a duration $D$ and is preceded by two priority tone slots ($PRS$), and a preamble ($P$). It is followed by a response inter-frame space ($RIFS$), the ACK, and finally, the contention inter-frame space ($CIFS$).
Thus, a successful transmission has a duration $T_s \doteq 2PRS+P+D+RIFS+ACK+CIFS.$ In the case of a collision, the stations
set the virtual carrier sense (VCS) timer equal to $EIFS,$ where $EIFS$ is the extended inter-frame space used by 1901, and then the channel state is \emph{idle}. Hence, a collision has a duration $T_c \doteq EIFS.$
Finally, we assume that all the packets use the same physical rate.

\begin{table}[!htb]
	\small
	\begin{center}
		\begin{tabular}{p{5cm}p{2cm}}
			Parameter & Duration ($\mu s$)\\
			\hline\hline
			Slot $\sigma$, Priority slot $PRS$ & 35.84 \\
			$CIFS$ & 100.00 \\
			$RIFS$ & 140.00 \\
			Preamble $P$, $ACK$   & 110.48 \\
			Frame duration $D$ & 2500.00 \\
			$EIFS$ & 2920.64 \\
			\hline
		\end{tabular}
	\end{center}
	\vspace{-0.1cm}
	\caption{Simulation parameters.}
	\vspace*{-0.4cm}
	\label{table:1901parslots}
\end{table}

\subsection{Experimental Validation of our Simulator}
\label{sec:experiments}
We use simulations to evaluate 1901 performance. To this end, we wrote a Matlab simulator that implements the full CSMA/CA mechanism of 1901.\footnote{Our simulator and the guidelines to reproduce all the testbed experiments of this work are available in~\cite{techreportexp}.}
In this subsection, we validate the accuracy of our model and simulator with experimental results from a HomePlug AV testbed. 

We built a testbed of 7 stations, each comprising a PLC interface.
The stations are ALIX boards running the OpenWrt Linux distribution~\cite{openwrt}. Each board is equipped with a Homeplug AV miniPCI card (Intellon INT6300 chip). In our experiments, $N$ stations send UDP traffic (at a rate higher than the link capacities) to the same non-transmitting station using \emph{iperf} {for 240s.} We run experiments for $1\le N \le 6.$ At the end of each test we request the number of collided and successfully transmitted frames from each station using the \emph{Qualcomm Atheros Open Powerline Toolkit}~\cite{openplc}.
Using this information, we evaluate the collision
probability. We compare the collision probability measured on the testbed with the one obtained with our model. To compute this probability from our model, we proceed as follows. Let $\gamma$ be the probability that a transmission in the system collides.
We use our model to compute the steady-state expected number of nodes $\hat{n}_0, \ldots, \hat{n}_{m-1}$ at each backoff stage. The probability that a given transmission in the system corresponds to a station at backoff stage $i$ is given by $\hat{n}_i\tau_i/\sum_{i=0}^{m-1}{\hat{n}_i\tau_i}$. We thus have
$
\gamma =\sum_{i=0}^{m-1}\hat{n}_i\tau_ip_i/\sum_{i=0}^{m-1}\hat{n}_i\tau_i
$.

\begin{figure}[!htb]
	\centering
	\includegraphics[scale=0.39]{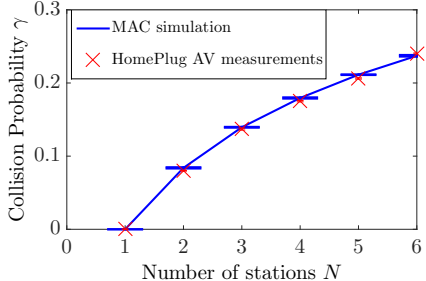}
	\caption{Collision probability obtained by simulation and experiments with HomePlug AV devices for the default class CA1 of 1901 given in Table~\ref{table:1901par}.
	}
	\label{fig:experiments}
\end{figure}

Figure~\ref{fig:experiments} shows the average collision probabilities, {along with confidence intervals,} obtained from 10 testbed experiments and 10 simulation runs ({note that confidence intervals are so small that they can barely be appreciated}). We observe an excellent fit between experimental 
and simulation results.

Contrary to some existing 802.11 interfaces, the MAC parameters of the HomePlug AV devices cannot be modified, as they are stored in the firmware, and the required offsets of their binary values are not publicly available. Thus, the following results have been obtained with our validated simulator.

%%%%%%%%%%%%%%%%%%%%%%%%%%%%%%%%
%%% 2. Model comparison: drift vs D.A. %%%
%%%%%%%%%%%%%%%%%%%%%%%%%%%%%%%%
\subsection{Comparison with Decoupling Assumption Model}
\begin{figure}[t]
	\centering
	\includegraphics[scale=0.39]{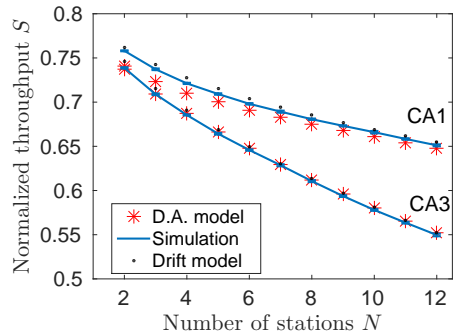}
	\caption{Throughput obtained by simulation, with our model, and the models based on the decoupling assumption (D.A.), for the default configurations of 1901 given in Table~\ref{table:1901par}.}
	\vspace*{-0.2cm}
	\label{fig:conf_1901}
\end{figure}

We now compare our model (which hereafter we refer to as ``drift model'') with the D.A. model for various configurations and number of stations.
In Figure~\ref{fig:conf_1901}, we show the throughput obtained by 1901 with the default parameters for the two priority classes CA1 and CA3 (CA0 and CA2 are equivalent).
We also show the throughput predicted by the two models.
The model based on the decoupling assumption is {substantially} less accurate for CA1 when $N$ is small, because the class CA1 uses larger contention windows, which increases the time spent in backoff and, as a result, the coupling between stations.

\begin{figure}[htb!]
	\centering
	\vspace{-0.2cm}
	\includegraphics[scale=0.39]{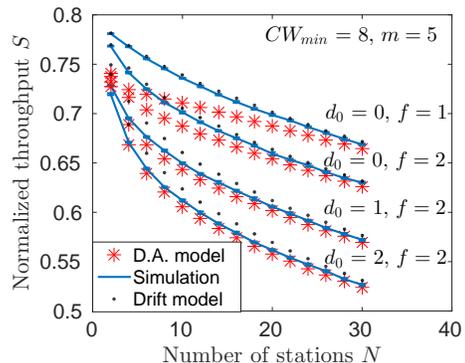}
	\vspace{-0.1cm}
	\caption{Throughput obtained by simulation, with the drift model, and the D.A. model for different configurations.
		The initial values $d_i$ of the deferral counter at each backoff stage are given by $d_i = f^i (d_0+1) - 1.$}
	\vspace{-0.1cm}
	\label{fig:conf_m4_cw8}
\end{figure}

We now study the accuracy of the two models in more general settings.
To this end, we introduce a factor $f$, such that at each stage $i$, the value of $d_i$ is given by $d_i = f^i(d_0 + 1) -1$.
This enables us to define various sequences of values for the $d_i$'s, using only $f$ and $d_0$. At each stage $i,$ $CW_i$ is given by $CW_i = 2^i CW_{min},$ and there are $m$ backoff stages ($i\in\{0,m-1\}$).
In Figure~\ref{fig:conf_m4_cw8}, we show the throughput for various such values of $d_0$ and $f$, with $CW_{min} = 8$ and $m=5$.
We observe that the D.A.\ model achieves good accuracy when the $d_i$'s are large, because in these configurations, the deferral counter is less likely to expire, which reduces the coupling among stations.
Note that the drift model achieves good accuracy when the $d_i$'s are small, while there is a small deviation for large $d_i$'s; this is due to {\ the assumptions of our network model (NM) of Section~\ref{sec:tr_analys}}, 
which are not used by the D.A.\ model.\footnote{The 802.11 model that does not rely on the decoupling assumption~\cite{sharma} has a similar deviation compared to Bianchi's model~\cite{bianchi}.}

\begin{figure}[htb!]
	\centering
	\vspace*{-0.25cm}
	\includegraphics[scale=0.39]{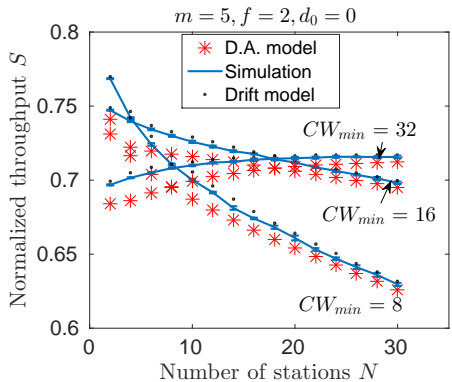}
	\vspace{-0.2cm}
	\caption{Throughput obtained by simulation, with the drift model, and the D.A. model for various values of $CW_{min}.$ }
	\vspace{-0.2cm}
	\label{fig:conf_dc0_f2}
\end{figure}

Finally, in Figure~\ref{fig:conf_dc0_f2} we show the throughput for different values for $CW_{min}$.
In all cases, the drift model closely follows simulation results, in contrast to the D.A.\ model. 

\subsection{Uniqueness of the Solution and Selected Counterexample}
\label{sec:transient_exp}
One of the fundamental results of our steady-state analysis is that there is a unique equilibrium for configurations that satisfy~\eqref{eq-cond}.
In this subsection we investigate the performance of the system depending on the (non-)unicity of the equilibrium of the dynamical system. To this end, we explore a counterexample of a configuration that does not satisfy~\eqref{eq-cond} and does not yield a unique equilibrium for the dynamical system~\eqref{eq-dynsys}.
An example of such a configuration, which yields 3 equilibrium points for $N=10$,\footnote{These 3 equilibrium points have been obtained by plotting $\Phi(p_e)$ (see the proof of Theorem~\ref{thm:unicity}) and computing the fixed-points for which $p_e = \Phi(p_e)$.} is the following:
\begin{equation}
\label{eq-conf-non-uniq}
\{CW_i,d_i\} = 
\begin{cases}
\{32,3\},\quad 0 \le i \le 3\\
\{4,\infty\},\quad \text{ }  4 \le i \le 53 \\ 
\{64,3\},\quad 54 \le i \le 59.
\end{cases}
\end{equation}

\begin{figure}[!htb]
	\centering
	\vspace{-0.2cm}
	\hspace*{-0.3cm}
	\includegraphics[width=0.25\textwidth]{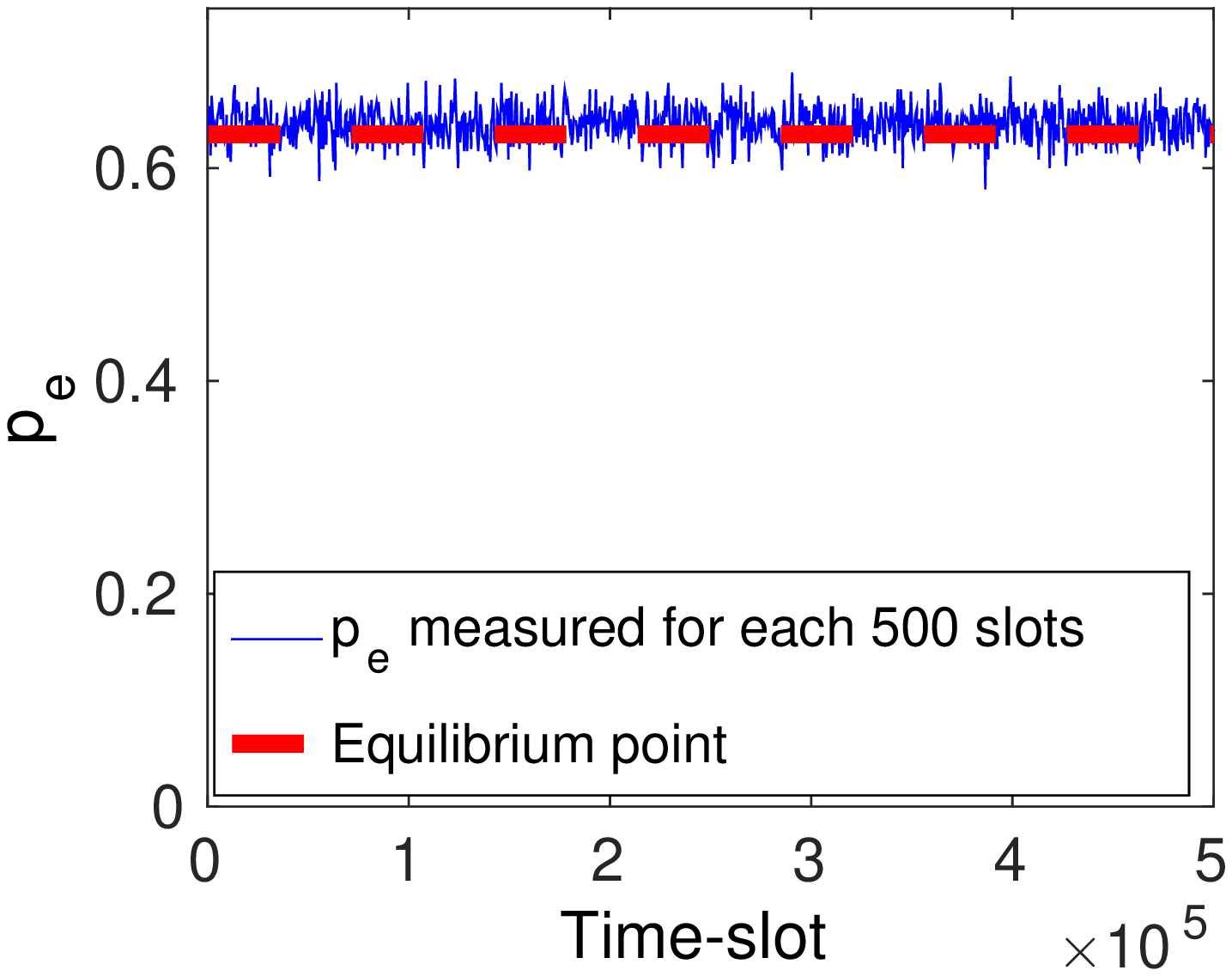}~\hspace*{-0.3cm}
	\includegraphics[width=0.25\textwidth]{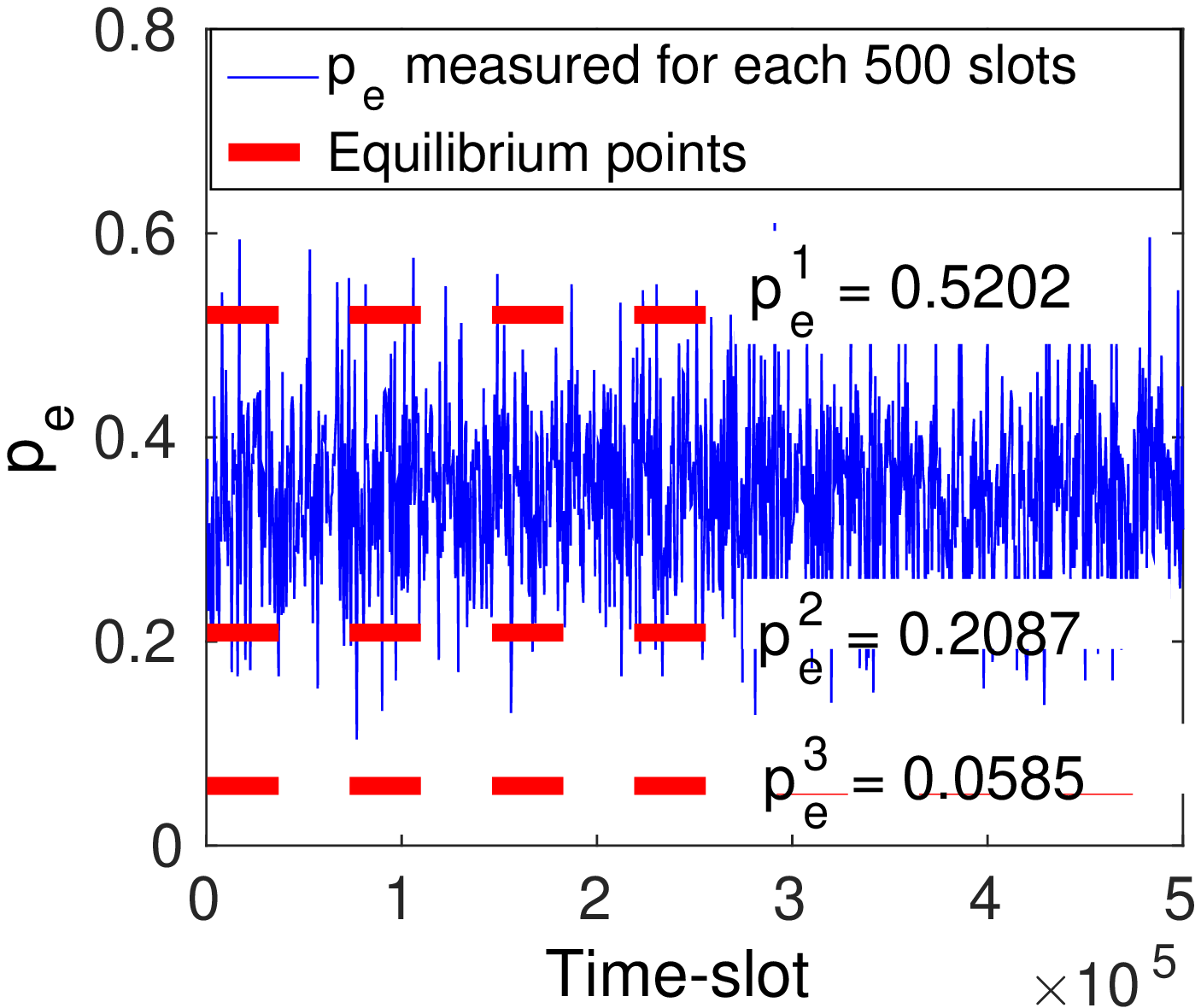}
	\vspace*{-0.55cm}
	\caption{Simulation of a system with a unique equilibrium point (left, configuration of CA1 class) and with 3 equilibrium points (right, configuration given by~\eqref{eq-conf-non-uniq}).
		The probability $p_e$ is computed for each 500 slots and is shown for one simulation run (plain black). The values of the equilibrium point(s) are also shown for each system (dashed red). The same behavior was observed for both systems for multiple simulation runs, not shown here.}
	\vspace*{-0.1cm}
	\label{fig:short_term_avg}
\end{figure}

To study this configuration, we compute the instantaneous $p_e,$ i.e, the probability that a time-slot is idle, for every 500 slots in simulation. Figure~\ref{fig:short_term_avg} shows the results for the CA1 class and for the configuration given by~\eqref{eq-conf-non-uniq}. We observe that for CA1 class, for which we have a unique equilibrium, the instantaneous $p_e$ is approximately equal to the one given by the equilibrium point of the dynamical system~\eqref{eq-dynsys}. However, for configuration~\eqref{eq-conf-non-uniq} $p_e$ oscillates between two of the equilibrium points and the value of $p_e$ averaged the entire simulation run is not equal to any of the equilibrium points; indeed the average $p_e$ obtained by one simulation run is 0.3478, whereas the values of the equilibrium points of~\eqref{eq-dynsys} are $(p_e^1,p_e^2,p_e^3) = (0.5202,0.2087,0.0585).$

Our results show that the equilibrium points~\eqref{eq-ni} are not sufficient to characterize the performance of the real system when~\eqref{eq-ni} are not unique: The real system might oscillate and, as a result, the behavior might not be close to any of the equilibrium points. They also suggest that such configurations should be avoided as they might lead to an unstable thus, undesirable behavior. {Indeed, multiple equilibria yield metastable regimes and typically involve severe unfairness or network collapse. For instance, with configuration~\eqref{eq-conf-non-uniq} some stations remain at a state with $CW_i = 4$ for long periods, leading to a very high collision probability and low throughput. The problems resulting from metastable regimes are reported in~\cite{altmanmetast} for 802.11.}

\subsection{Accuracy of the Drift Model in the Transient Regime}
The above experiments have focused on the accuracy of our steady-state analysis for the stationary regime. In the following, we investigate the accuracy of the analysis for the transient regime. To this end, we consider a system with $N=20$ stations and two different configurations, and compare the expected number of stations $\bar{\mathbf{n}}(t)$ obtained from~\eqref{eq-dynsys} and from simulations, as a function of the time slot $t,$ when the initial condition at time slot 0 is $\mathbf{n}(0)= \{20, 0, 0, ..., 0, 0\}.$ 

\begin{figure}[!htb]
	\centering
	\vspace{-0.1cm}
	\includegraphics[width=0.2\textwidth]{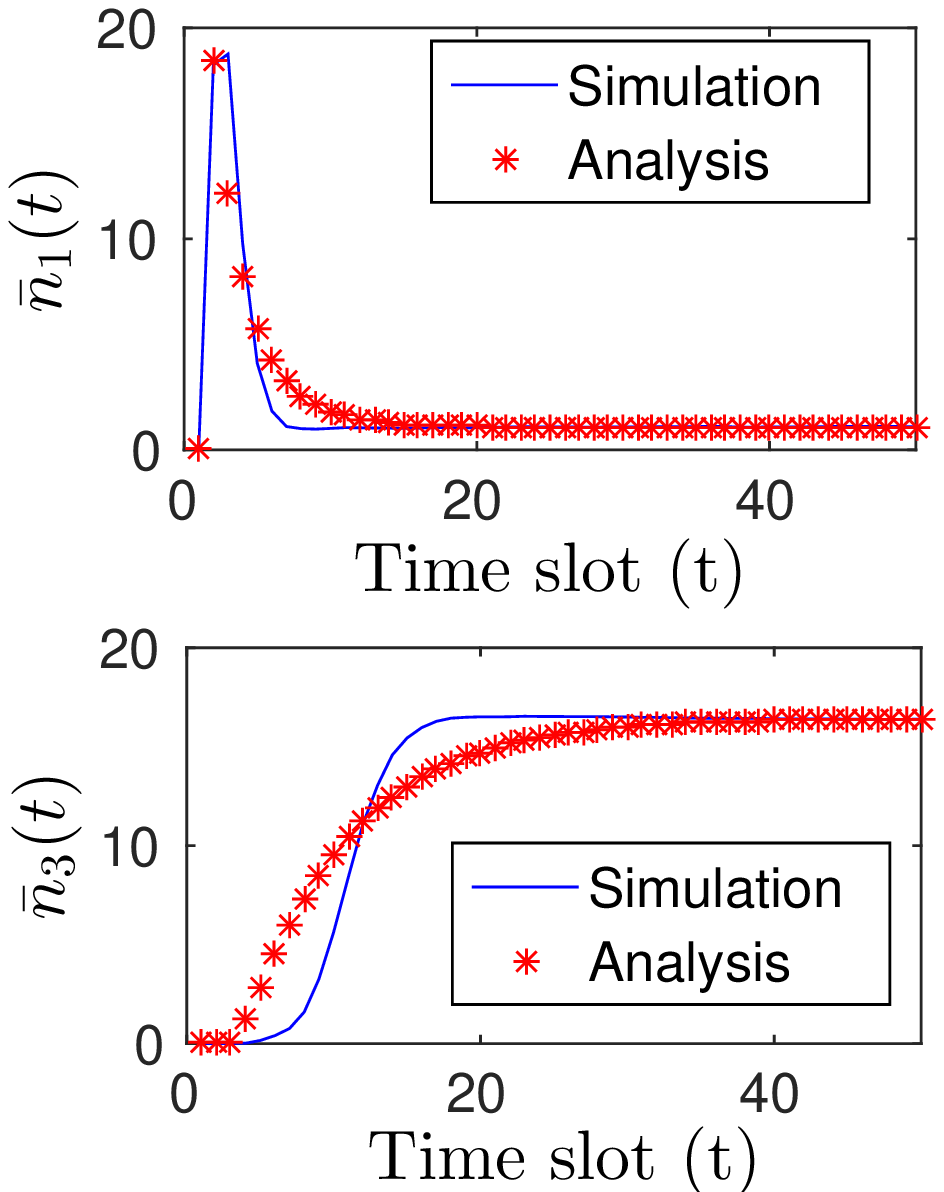}~\hspace*{-0.3cm}
	\includegraphics[width=0.2\textwidth]{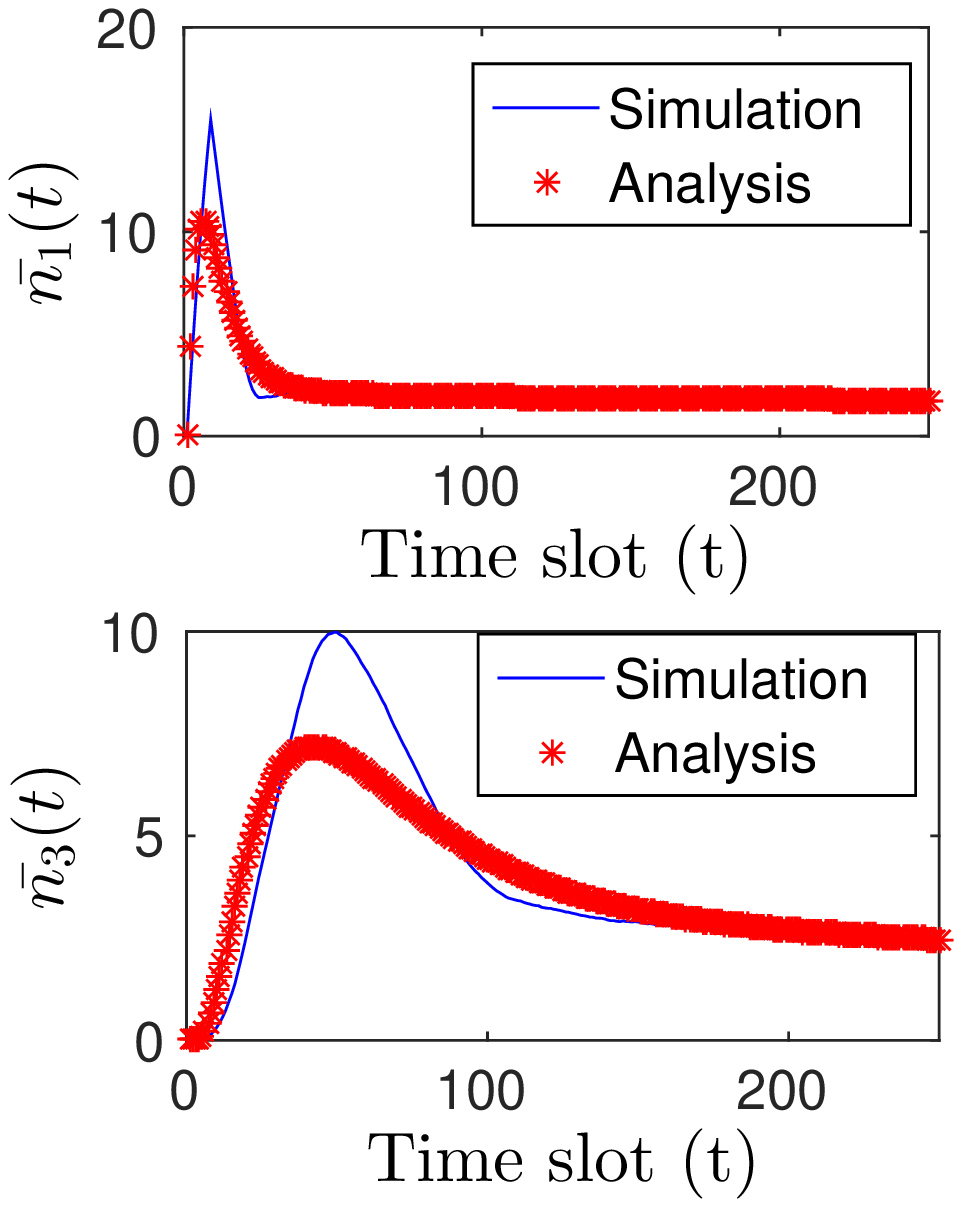}
	\vspace*{-0.25cm}
	\caption{Convergence to the equilibrium point of the number of stations for backoff stages 1 and 3, for the CA1 class (left) and for a configuration with $d_i \to \infty \text{ and } CW_i = 2^{i}CW_0,\,\forall i,$ $CW_0 = 8,\, m=7,$  (right).
		Both the expected values obtained from~\eqref{eq-dynsys} and the average values obtained from 2000 simulation runs are shown.}
	\vspace{-0.1cm}
	\label{fig:convergence}
\end{figure}

We focus on a configuration that follows the 1901 standard, i.e., CA1 class, and on a configuration that follows the 802.11 standard, i.e., the deferral counter does not expire. 
The results from the experiments described above are shown in Figure~\ref{fig:convergence}.
We observe that our model works well both in terms of accuracy and of convergence times. As far as accuracy is concerned,  
there is slightly higher inaccuracy in the transient regime than in the stationary regime, which is due to the assumption on the constant transition probabilities $\beta_i$ and $\tau_i$.\footnote{To confirm that the deviations are due to this assumption, we simulated the Markov chain $\mathbf{X}(t)$ with constant transition probabilities, and verified that the trajectory of $\mathbf{X}(t)$ averaged over 300 runs coincides with the~solution~of~\eqref{eq-dynsys}.}
The convergence time to the equilibrium points is also captured by our model with reasonable accuracy. This time is higher for the 802.11 system for two reasons: 1) in the 802.11 system, the stations are allowed to have larger backoff counters and to move into higher backoff stages; 2) in the 1901 system, the stations change their backoff stage with a higher probability than in 802.11 due to the deferral counter.

\subsection{Configuration Guidelines with Respect to~\eqref{eq-cond}}

\begin{figure}[!htb]
	\centering\vspace*{-0.2cm}
	\hspace*{-3pt}\includegraphics[width=0.255\textwidth]{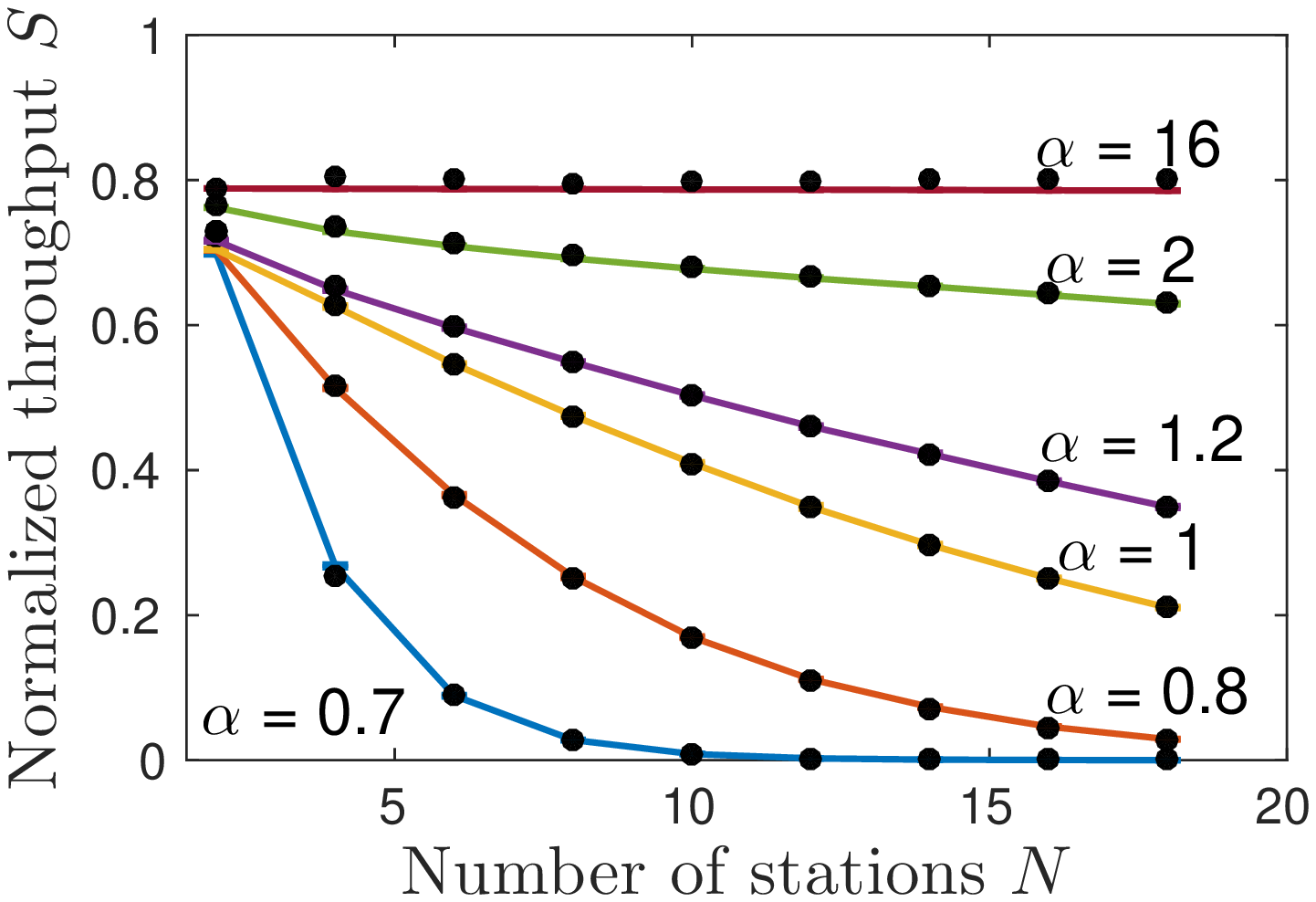}~\hspace*{-6pt}
	\includegraphics[width=0.255\textwidth]{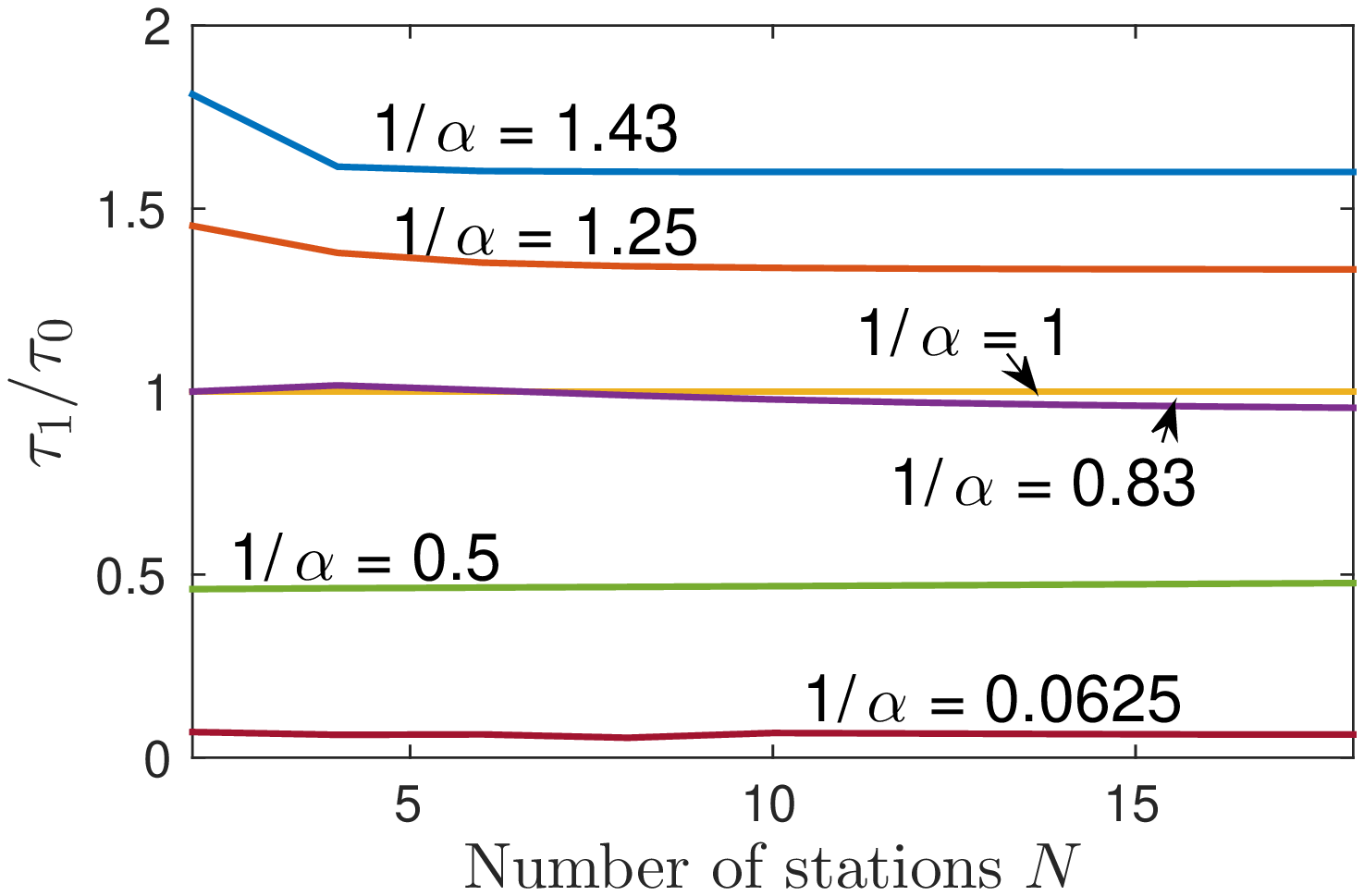}
	\vspace*{-0.5cm}
	\caption{Performance of 1901 with parameters $CW_i = \lfloor \alpha^{i-1} \cdot 8\rfloor,$ and $  d_i = \lceil \alpha^{i-1} -1 \rceil$ for different values of $\alpha.$ Lines represent throughput obtained by simulation and points show throughput computed by our model (left). We also present the ratio $\tau_1/\tau_0$ computed using our model (right).}
	\vspace{-0.2cm}
	\label{fig:tau_i_guid}
\end{figure}

As discussed in Section~\ref{sec:uniqueness},~\eqref{eq-cond} is not only a condition for uniqueness, but also a configuration guideline for proper reaction to high contention. Jumping to the next backoff stage is an indication of high contention hence, to dissolve the current contention, the transmission aggressiveness should decrease, that is $\tau_{i+1} < \tau_i.$  We now show that configurations where $\tau_i$ is increasing with $i$ perform poorly.
To confirm this, in the following we run several experiments with different $\alpha$ values, where $\alpha$ is the multiplicative factor of the contention windows between successive backoff stages, i.e., $CW_{i+1} = \alpha CW_i.$
Figure~\ref{fig:tau_i_guid} presents throughput obtained by simulation and with our model for various values of $\alpha,$ with $CW_i = \lfloor \alpha^{i-1} \cdot 8\rfloor,$ $  d_i = \lceil \alpha^{i-1} -1 \rceil,\, 0\le i \le 4.$ Results show that configurations with $\tau_i$ increasing, i.e., $\alpha <1$, yield poor performance. This {supports} our argument that~\eqref{eq-cond} should be met to ensure good performance. Theorem~\ref{thm:taui_dec} provides some configuration guidelines to ensure that~\eqref{eq-cond} is satisfied.

{
	As it can be seen from the figure, throughput performance improves for large $\alpha$. However, a closer look at the protocol behavior for different $\alpha$'s reveals that, while large $\alpha$'s provide very good throughput performance, they also suffer from severe unfairness. Indeed, for such configurations only one station grasps the channel, while the others move to higher backoff stages with much larger $CW_i$ values and barely transmit. This shows that throughput considerations are not sufficient to properly evaluate the suitability of a given 1901 configuration, and short-term fairness also needs to be taken into account.}

\section{Conclusion}
\label{sec:conclusion}
Although the IEEE 1901 CSMA/CA protocol is adopted by the vast majority of PLC devices nowadays, it has received little attention from the research community so far. In this paper, we focus on the performance analysis of this protocol. Our analysis comprises performance in steady-state as well as in the transient regime, and involves both the long-term and the short-term dynamics of 1901. One of the key results of the analysis is the finding that the decoupling assumption, which is commonly adopted for the analysis of MAC protocols such as IEEE 802.11 and 1901, might not hold for 1901. This is due to the coupling that 1901 introduces to the stations contending for the medium. Building on this finding, we have proposed a model that does not rely on the decoupling assumption, and as a result, substantially improves the accuracy of previous analyses. Accuracy is particularly improved for networks with a small number of stations, which is the most frequent scenario in practice. We have shown that our model admits a unique solution for a wide range of configurations.

\section*{Appendix}
Table~\ref{table:summ} can be used as a quick reference for the equations of Section~\ref{sec:analysis}.
\begin{table}[!htb]
	\vspace{-0.2em}
	\begin{center}
		\footnotesize
		\renewcommand{\arraystretch}{2.8}
		\begin{tabular}{p{1cm} p{6.5cm} p{0.3cm}}
			Notation & Value & Eq. \\
			\hline
			\hline
			$p_i$ & $1-\frac{p_e}{1-\tau_i}$ & \eqref{eq-pi} \\
			$x_k^i$ &  $\sum_{j = d_i+1}^{k} \binom{k}{j} p_i^{j} (1-p_i)^{k-j}$ & \eqref{eq:def_xk}\\
			$p_e$ & $\prod_{k = 0}^{m-1}{(1-\tau_k)^{\bar{n}_k}}$ & -\\
			$bc_i$ & $\frac{\sum_{k = d_i+1}^{CW_i-1}{\left[ (k+1)(1-x^i_{k})+\sum_{j=d_i+1}^{k}{j(x^i_j-x^i_{j-1})} \right]}}{CW_i}$ & \eqref{eq-bc_i}\\
			& \quad $+\frac{(d_i+1)(d_i+2)}{2 CW_i}$ &\\
			$t_i$ & $ \sum_{k = d_i+1}^{CW_i-1}{\frac{1}{CW_i}(1-x^i_{k})}+\frac{d_i+1}{CW_i}$ & \eqref{eq:tn}\\
			$\tau_i$ & $\frac{\sum_{k = d_i+1}^{CW_i-1}{\frac{1}{CW_i}(1-x^i_{k})}+\frac{d_i+1}{CW_i}}{bc_i}$ & \eqref{eq-tau_i}\\
			$\beta_i$ & $\frac{\sum_{k =d_i+1}^{CW_i-1}{\frac{1}{CW_i}\sum_{j=d_i+1}^{k}{(x^i_j-x^i_{j-1})}}}{bc_i}$ & \eqref{eq-beta_i}\\
			$B_i$ & $\frac{\frac{CW_{i}(CW_{i}-1)}{2}-\sum_{k = d_{i}+1}^{CW_{i}-1}{(CW_i-1-k){x^i_k}}}{CW_{i} - \sum_{k=d_{i}+1}^{CW_{i}-1}{x^i_k}}$ & \eqref{eq-Bi-simple}\\
			$K_i$ & $K_0= 1, K_i = \frac{\tau_{i-1} p_{i-1} + \beta_{i-1}}{\tau_i + \beta_i}, 1\leq i \leq m-2$ & \eqref{eq-ki} \\
			& $K_{m-1} = \frac{\tau_{m-2} p_{m-2} + \beta_{m-2}}{\tau_{m-1}(1-p_{m-1})} $ & \\
			$\bar{n}_i$ & $\bar{n}_0 =\frac{N}{\sum_{k = 0}^{m-1}{\prod_{j = 0}^{k}{K_j}}} ,  $ & \eqref{eq-ni}\\
			& $\bar{n}_i = \frac{N\prod_{j = 0}^{i}{K_j}}{\sum_{k = 0}^{m-1}{\prod_{j = 0}^{k}{K_j}}}  \, 1\leq i \leq m-1$ &\\
		\end{tabular}
	\end{center}
	\caption{Summary of Variables}
	\label{table:summ}
	\label{table:appendix}
\end{table}

\begin{lm} $B_i$ is an increasing function of $p_i,$ and $\tau_i$ is a decreasing function of $p_i$ for any $0\le i \le m-1$.
	\label{lemma:tau_bci}
\end{lm}
\begin{IEEEproof}
	The probability $\tau_i$ given by (\ref{eq-tau_i}) can be recast as
	\begin{equation}
	\tau_i = \frac{1}{B_i + 1}
	\end{equation}
	where $B_i$ is the expected number of backoff slots between two transmission attempts of a station that always stays at backoff stage $i$. 
	$B_i$ can be computed recursively and similarly to $bc_i$ in~\eqref{eq-bc_i}, and it is given by
	\begin{align}
	\label{eq-B1}
	B_i  = & \frac{d_i(d_i+1)}{2 CW_i}\\\nonumber
	& +\sum_{j = d_i+1}^{CW_i-1}{\frac{j(1-x^i_{j})+\sum_{k=d_i+1}^{j}{(k+B_i)(x^i_k-x^i_{k-1})}}{CW_i}}.
	\end{align}
	
	We simplify~\eqref{eq-B1} as
	{\small
		\begin{align}
		B_i =& \underbrace{\frac{d_i(d_i+1)}{2 CW_i} + \sum_{j = d_i+1}^{CW_i-1}\frac{j}{CW_i}}_{1}-\underbrace{ \sum_{j = d_i+1}^{CW_i-1}\frac{jx^i_j}{CW_i}}_{3}\nonumber\\
		&+\frac{1}{CW_i}\sum_{j = d_i+1}^{CW_i-1}\sum_{k=d_i+1}^{j}\left({ \underbrace{B_i(x^i_k-x^i_{k-1})}_{2}+ \underbrace{{k(x^i_k-x^i_{k-1})}}_3}\right)\nonumber\\
		= & \underbrace{\frac{CW_i-1}{2}}_{1} + \frac{1}{CW_i}\sum_{j = d_i+1}^{CW_i-1}\left( \underbrace{B_i x^i_{j}}_{2}\underbrace{-\sum_{k=d_i+1}^{j-1}{x^i_k}}_{3} \right),
		\label{eq-B}
		\end{align}
	}
	where we have used that $x^i_{d_i} = 0,$ for all $p_i\in[0,1]$ by the definition of $x^i_k$, and we have combined the terms with the same under-brace indexes. 
	
	To prove the lemma, we proceed as follows. ($i$) First, we compute $d B_i/d p_i.$ ($ii$) Second, we show that this derivative is positive at $p_i = 1.$
	($iii$) Third, we show that if the derivative is negative for some $0 < p_i^* < 1,$ it will also be negative at any value $p_i > p_i^*.$
	The proof then follows by contradiction: if the derivative was negative at some $p_i^*$, it would also be negative at $p_i = 1$, which would contradict our previous result.
	
	($i$) 
	The derivative of $B_i$ can be computed as
	\begin{equation}
	\frac{d B_i}{d p_i} = \sum_{k = d_i+1}^{CW_i-1}{\frac{\partial B_i}{\partial x^i_k}\frac{d x^i_k}{d p_i}}.
	\end{equation}
	The partial derivative $\partial B_i/\partial x^i_k$ can be computed by~(\ref{eq-B})~as
	\begin{equation}
	\label{eq-partialB}
	\frac{\partial B_i}{\partial x^i_k} = \frac{B_i - (CW_i-1-k)}{CW_i} + \frac{\partial B_i}{\partial x^i_k}\sum_{j = d_i+1}^{CW_i-1}{\frac{x^i_{j}}{CW_i}},
	\end{equation}
	which yields 
	\begin{align}
	\label{eq-partialB2}
	\frac{d B_i}{d p_i} =
	\frac{\sum_{k = d_i+1}^{CW_i-1}{(B_i - (CW_i-1-k))\frac{d x^i_k}{d p_i}}}{CW_i-\sum_{j = d_i+1}^{CW_i-1}{x^i_{j}}}.
	\end{align}
	
	To compute $d x^i_k/d p_i$, we observe that $x^i_k$ is the complementary cumulative function of a binomial distribution and it can be expressed as the incomplete beta function $x^i_k = 1 - I_{1-p_i}(k-d_i,d_i+1)$~\cite{binom}.
	The derivative of an incomplete beta function is computed in~\cite{betainc}, and by using this we obtain 
	\begin{equation}
	\label{eq-partialkx}
	\frac{d x^i_k}{d p_i} = \frac{k!}{(k-d_i-1)!d_i!}p_i^{d_i}(1-p_i)^{k-d_i-1}.
	\end{equation}
	
	($ii$) Next, we show that $d B_i/d p_i > 0$ at $p_i = 1.$
	Note that $x^i_k=1$ at $p_i=1$ for all $d_i+1\le k \le CW_i -1.$ 
	Given this, we have
	\begin{equation}
	B_i =  \frac{d_i(d_i+1)}{2 CW_i} + \frac{CW_i - d_i -1}{CW_i}(d_i+1+B_i).
	\label{Bi_at1}
	\end{equation}
	
	Solving~\eqref{Bi_at1} over $B_i$ yields $B_i =  CW_i-d_i/2 -1$ at $p_i = 1.$
	Now, notice that $d x^i_k/d p_i = 0$ at $p_i=1$ for all $d_i+1 < k \le CW_i -1,$ and $d x^i_{d_i+1}/d p_i = d_i+1$ from~\eqref{eq-partialkx}.  Substituting in (\ref{eq-partialB2}) yields $d B_i/d p_i = d_i/2 +1,$ i.e., $d B_i/d p_i > 0$ at~$p_i = 1.$
	
	$iii$) Next, we show that if $d B_i/d p_i$ was negative at some value $p_i^*,$ then it would also be negative for any $p_i>p_i^*.$ Observe that in~\eqref{eq-partialB2} some terms are negative for $k< CW_i-1 - B_i.$ 
	Let us assume that the derivative is negative at $p_i^*.$ 
	Let $l = \lceil CW_i - 1 - B_i(p_i^*) \rceil$. Given (\ref{eq-partialB2}),
	we can express $d B_i/d p_i$ as the product of two terms, $d B_i/d p_i = f_1(p_i) f_2(p_i)$, where
	\begin{equation}\nonumber
	f_1(p_i) \doteq \frac{d x_l^i /d p_i}{CW_i - \sum_{j = d_i+1}^{CW_i-1}x_j^i},
	\end{equation}
	\begin{equation}\nonumber
	f_2(p_i) \doteq \sum_{k = d_i +1}^{CW_i-1}(B_i-(CW_i - 1 -k)) \frac{d x_k^i/d p_i}{d x_l^i/d p_i}.
	\end{equation}
	
	Note that $f_1(p_i) > 0 \ \forall p_i$, which implies $d B_i/d p_i < 0$ if and only if $f_2(p_i) < 0$. Note also that
	\begin{align}\label{eq-df2}
	\frac{d f_2(p_i)}{d p_i} = & \sum_{k = d_i+1}^{CW_i-1}\frac{d B_i}{d p_i} \frac{d x_k^i / d p_i}{d x_k^l / d p_i} \\\nonumber
	& + \sum_{k = d_i+1}^{l-1}(B_i-(CW_i-1-k)) \frac{d}{d p_i}\left(\frac{d x_k^i / d p_i}{d x_l^i / d p_i}\right) \\\nonumber
	& + \sum_{k = l+1}^{CW_i-1}(B_i-(CW_i-1-k)) \frac{d}{d p_i}\left(\frac{d x_k^i / d p_i}{d x_l^i / d p_i}\right)
	\end{align}
	and
	\begin{equation}\nonumber
	\frac{d}{d p_i}\left(\frac{d x_k^i / d p_i}{d x_l^i / d p_i}\right) = -\frac{k!(l-d_i-1)!}{l!(k-d_i-1)!} (k-l)(1-p_i)^{k-l-1},
	\end{equation}
	which is positive for $k < l$ and negative for $k > l$. From the above equations, it follows that as long as $CW_i-1-(l-1) > B_i(p_i) > CW_i-1-l$ and $d B_i/d p_i < 0$, we have $d f_2/d p_i < 0$. 
	
	Building on the above, next we show that $B_i(p_i)$ decreases for $p_i \in [p_i^*, p_i^l]$, where $p_i^l$ is the $p_i$ value for which $B_i(p_i^l) -(CW_i -1 -l) = 0$. At $p_i = p_i^*$ we have $f_2(p_i^*) < 0$, $d B_i/dp_i < 0$ and $d f_2/dp_i < 0$. Let us assume that, before $B_i(p_i)$ decreases down to $CW_i-1-l$, there is some $\hat{p_i} > p_i^*$ for which $d B_i/ d p_i \geq 0$. This implies that for some $p_i' \in (p_i^*,\hat{p_i})$, $f_2(p_i)$ has to stop decreasing, i.e., $d f_2(p_i')/d p_i = 0$. Note that, since $f_2(p_i)$ decreases in $[p_i^*,p_i']$, we have $f_2(p_i) < 0$ for $p_i \in [p_i^*,p_i']$, which implies that $B_i(p_i)$ decreases in $[p_i^*,p_i']$. Also, since $CW_i-1-(l-1) > B_i(p_i^*) > CW_i-1-l$ and (by assumption) $B_i(p_i)$ does not reach $CW_i-1-l$, we also have $CW_i-1-(l-1) > B_i(p_i') > CW_i-1-l$. However, we have seen that $f_2(p_i') < 0$ and $CW_i-1-(l-1) > B_i(p_i') > CW_i-1-l$ implies $d f_2(p_i')/d p_i < 0$, which contradicts $d f_2(p_i')/d p_i = 0$. Hence, our initial assumption does not hold, which implies that $d B_i/d p_i$ decreases until $B_i$ reaches $CW_i-1-l$, i.e., $d B_i/d p_i < 0$ for $p_i \in [p_i^*, p_i^l]$.
	
	%. Since $f_2(p_i^*) < 0$ and $d f_2(p_i)/d p_i \leq 0$ for $p_i \in [p_i^*,p_i']$, this implies
	
	Following the same rationale for $p_i \in [p_i^l, p_i^{l+1}]$ and choosing
	\begin{equation}\nonumber
	f_1(p_i) = \frac{d x_{l+1}^i /d p_i}{CW_i - \sum_{j = d_i+1}^{CW_i-1}x_j^i},
	\end{equation}
	\begin{equation}\nonumber
	f_2(p_i) = \sum_{k = d_i +1}^{CW_i-1}(B_i-(CW_i - 1 -k)) \frac{d x_k^i/d p_i}{d x_{l+1}^i/d p_i},
	\end{equation}
	we can prove that $d B_i/d p_i < 0$ for $p_i \in [p_i^l, p_i^{l+1}]$. We can repeat this recursively to show that $d B_i/d p_i < 0$ for $p_i \in [p_i^{l+1}, p_i^{l+2}]$, $p_i \in [p_i^{l+2}, p_i^{l+3}]$ until reaching $p_i \in [p_i^{CW_i-2}, p_i^{CW_i-1}]$. If $1 < p_i^{CW_i-1}$, the proof is completed. Otherwise, we can follow a similar argument to show that $d B_i/d p_i < 0$ for $p_i \in [p_i^{CW_i-1}, 1]$. Indeed, let us assume that $f_2(p_i) \geq 0$ for some $\hat{p_i} \in [p_i^{CW_i-1},1]$. This implies that for some $p_i' \in  [p_i^{CW_i-1},\hat{p_i}],$ we have $d f_2(p_i')/d p_i = 0$ and $f_2(p_i') < 0$, which is not possible according to (\ref{eq-df2}).
	From the above, if $d B_i/d p_i$ was negative at any $p_i^*$, it would also be negative for all $p_i > p_i^*$. Since this contradicts result ($ii$), we conclude that $d B_i/d p_i \geq 0$ for $p_i \in [0,1]$.
\end{IEEEproof}

\begin{cor}
	$\beta_i$ is an increasing function of $p_i.$
	\label{cor:betai}
\end{cor}
\begin{IEEEproof}
	From~\eqref{eq-tau_i} and~\eqref{eq-beta_i}, we have $\beta_i = 1/bc_i-\tau_i.$
	$bc_i$ in~\eqref{eq-bc_i} can be simplified as follows:
	\begin{equation}
	bc_i = \frac{CW_i+1}{2}-\frac{\sum_{k = d_i+1}^{CW_i-1}{\sum_{j=d_i+1}^{k}x^i_j}}{CW_i}
	\label{eq-bci_s}.
	\end{equation}
	Since $d x^i_k/d p_i > 0, k \ge d_i+1,$ $bc_i$ is decreasing with $p_i.$ 
	Hence, by using Lemma~\ref{lemma:tau_bci}, $\beta_i$ is increasing with $p_i.$
\end{IEEEproof}

\begin{cor}
	\label{cor:bi-der}
	For any value of $i,$ {$d B_i/d p_i < B_i^2{/(1-p_i)}$  $\forall p_i\in [0,1)$, and $d B_i/d p_i < B_i{/p_i}$ $\forall p_i\in (0,1].$}
\end{cor}
\begin{IEEEproof}
	%  We distinguish two cases to prove the corollary for $pi\in [0,1)$, one for $d_i = 0,$ and one for $d_i \neq 0.$
	We~start~with~the~first~\mbox{inequality.} The derivative $d B_i/d p_i$ can be computed from~(\ref{eq-Bi-simple})~as
	\begin{align}
	\label{eq-partialBineq}
	\frac{d B_i}{d p_i} =
	\frac{\sum_{k = d_i+1}^{CW_i-1}{(B_i - (CW_i-1-k))\frac{d x^i_k}{d p_i}}}{CW_i-\sum_{j = d_i+1}^{CW_i-1}{x^i_{j}}}.
	\end{align}

	To prove the lemma, we distinguish two cases: one for $d_i=0$ and one for $d_i > 0.$
	First, let us study~\eqref{eq-partialBineq} with $d_i = 0.$ 
	We have $d x_k^i/d p_i = k (1-p_i)^{k-1}$ and $ x_k^i = 1-(1-p_i)^{k}.$ 
	Let $h(p_i)=\sum_{k = 0}^{CW_i-1}{k (1-p_i)^{k}}/\sum_{k = 0}^{CW_i-1}{(1-p_i)^{k}}.$ We now show that $h$ decreases with $p_i.$ Let also $G(p_i) =\sum_{k = 0}^{CW_i-1}{p_i^{k}} /\sum_{k = 0}^{CW_i-1}{k p_i^{k}}.$ By Lemma 5.1 in~\cite{kumar}, $G(p_i)$ is strictly decreasing with $p_i$ in $[0,1].$ Thus, $h(p_i) = 1/G(1-p_i)$ is also strictly decreasing with $p_i$ in $[0,1],$ and $h(p_i)\le h(0)=(CW_i-1)/2.$ Also, $B_i(p_i)\ge (CW_i-1)/2$ by Lemma~\ref{lemma:tau_bci}.
	Given the above, we have $h(p_i) \le B_i(p_i)$ and~\eqref{eq-partialBineq} yields
	\begin{align}
	\frac{d B_i}{d p_i} < \frac{B_i}{1-p_i} \frac{\sum_{k = 0}^{CW_i-1}{k (1-p_i)^{k}}}{\sum_{k = 0}^{CW_i-1}{(1-p_i)^{k}}}\leq \frac{B_i^2}{1-p_i}.\nonumber
	\end{align}
	% Thus,~\eqref{eq-condition} holds for $d_i=0$ and $p_i \ne 1.$
	% For $p_i=1,$ $d B_i/d p_i =1$ and clearly~\eqref{eq-condition} is true if $CW_i > 3 .$
	
	We now move to the case $d_i > 0.$
	{From~(\ref{eq-partialkx}) and~\eqref{eq:def_xk}, we have $d x^i_k/d p_i = k (x^i_k - x^i_{k-1}) /p_i.$ Thus,~\eqref{eq-partialBineq} yields
		\begin{equation}
		\frac{d B_i}{d p_i} < \frac{B_i}{p_i} \frac{ CW_i x^i_{CW_i-1} - \sum_{k = d_i+1}^{CW_i-1}{x^i_{k}}}{CW_i-\sum_{k = d_i+1}^{CW_i-1}{x^i_{k}}}\leq \frac{B_i}{p_i} x^i_{CW_i-1}.
		\label{eq-B_i-der-ineq}
		\end{equation}}
	From the above, we have
	\begin{align*}
	&	\frac{ x^i_{CW_i-1}}{p_i} \\
	& \quad	{=} \sum_{j = d_i+1}^{CW_i-1} \binom{CW_i-1}{j} p_i^{j-1} (1-p_i)^{CW_i-1-j} \nonumber\\
	&\quad	{=} \sum_{j = d_i+1}^{CW_i-1} \frac{CW_i-j}{j(1-p_i)} \binom{CW_i-1}{j-1} p_i^{j-1} (1-p_i)^{CW_i-1-(j-1)} \nonumber\\
	&\quad	{\leq}  \frac{CW_i-1}{2(1-p_i)} \sum_{j = d_i+1}^{CW_i-1} \binom{CW_i-1}{j-1} p_i^{j-1} (1-p_i)^{CW_i-1-(j-1)}\nonumber\\
	& \quad	{\leq}  \frac{CW_i-1}{2(1-p_i)} \leq \frac{B_i}{1-p_i}\nonumber.
	\end{align*}
	
	Thus, combining the above two equations we have { $d B_i/d p_i < B_i^2{/(1-p_i)}.$} Then, {$d B_i/d p_i < B_i{/p_i}$} follows from~\eqref{eq-B_i-der-ineq}, since $x^i_{CW_i-1} \le 1,$ which completes the proof.
\end{IEEEproof}

\begin{lm}
	Let us consider the expression of $\tau_i$ as a function of $p_e$ resulting from combining (\ref{eq-tau_i}) with~(\ref{eq-pi}). According to this expression, $\tau_i$ is an increasing function of $p_e.$
	\label{lemma:taui_vs_pe}
\end{lm}
\begin{IEEEproof}
	Since $\tau_i = 1/(B_i+1)$, we need to show that $d B_i/d p_e < 0$. Note that
	\begin{equation}{
		\frac{d B_i}{d p_e} = \frac{d B_i}{d p_i}\frac{d p_i}{d p_e}.}
	\label{eq-Bipe}
	\end{equation}
	
	From $p_i = 1 - p_e/(1-\tau_i)=1-p_e (B_i+1)/B_i$, we have
	\begin{align}{\
		\frac{d p_i}{d p_e} = - \frac{B_i+1}{B_i} + \frac{p_e}{B_i^2}\frac{d B_i}{d p_e}.}
	\label{eq-pipe}
	\end{align}
	
	Combining (\ref{eq-Bipe}) and (\ref{eq-pipe}) yields
	\begin{equation}{\
		\frac{d B_i}{d p_e} = - \frac{d B_i}{d p_i} \frac{B_i+1}{B_i} \frac{1}{1- \frac{p_e}{B_i^2}\frac{d B_i}{d p_i}}.}
	\label{eq-pe-der}
	\end{equation}
	
	Let us distinguish two cases to prove this lemma, one for $p_i = 1$ and the other for $0\le p_i <1.$
	First, for $p_i = 1,$ we have $p_e = 0$ from \eqref{eq-pi}. Thus,~\eqref{eq-pe-der} is smaller than 0.
	
	We now look at the case $0\le p_i <1.$
	From Lemma~\ref{lemma:tau_bci}, we have {$d B_i/d p_i > 0$} therefore, {$d B_i/d p_e < 0$} as long as
	\begin{equation}{
		\frac{d B_i}{d p_i} < \frac{B_i^2}{p_e} = \frac{B_i(B_i+1)}{1-p_i}.}
	\label{eq-condition}
	\end{equation}

	According to Corollary~\ref{cor:bi-der}, {$d B_i/d p_i < B_i^2{/(1-p_i)},$} which is a sufficient condition for~\eqref{eq-condition}. 
	This terminates the proof. 
\end{IEEEproof}

\begin{lm}\label{lm-ji} Let $j_i(p_i) \doteq \tau_i(p_i)p_i + \beta(p_i)$. Then $d j_i/d p_i \leq 1$.
\end{lm}
\begin{IEEEproof}	Let us define $J_i \doteq 1/j_i$. From~\eqref{eq:tn}, {\eqref{eq-tau_i}} and {\eqref{eq-beta_i}}, it can be seen that $\tau_i = t_i / bc_i,$ $\beta_i = (1 - t_i) / bc_i,$ hence $j_i = (1 - t_i (1 - p_i) ) /  bc_i$ which yields
	\begin{align}\label{eq-ji}
	J_i = \frac{\sum_{k = 0}^{CW_i-1}{\Big((k+1)(1-x_k^i)} +\sum_{j=0}^{k}{j(x_j^i-x_{j-1}^i)}\Big)}{CW_i - \sum_{k = 0}^{CW_i-1}{(1-x_k^i)(1-p_i)}}.
	\end{align}
	To simplify the exposition in this lemma, observe that in the above equation, we have all summations starting at $k=0,$ which means that in the following we set $x^i_k = 0$ and $d x^i_k / d p_i = 0$  for all $0\le k \le d_i.$
	
	An equivalent expression for $J_i$ is as follows
	\begin{align}\label{eq-ji2}
	J_i = &\ \frac{1}{CW_i}\sum_{k = 0}^{CW_i-1}{(1-x_k^i)(1-p_i)J_i} + \frac{CW_i-1}{2}\nonumber\\
	& - \frac{1}{CW_i}\sum_{k = 0}^{CW_i-1}{\sum_{j=0}^{k}{x_j^i}}.
	\end{align}
	
	By derivation of the above expression, we obtain
	\begin{align*}
	\frac{d J_i}{d p_i} = &\ \frac{1}{CW_i}\sum_{k = 0}^{CW_i-1}{(1-x_k^i)(1-p_i)\frac{d J_i}{d p_i}} \nonumber\\
	& - \frac{1}{CW_i}J_i\sum_{k = 0}^{CW_i-1}{\left(\frac{d x_k^i}{d p_i}(1-p_i)+(1-x_k^i)\right)}\nonumber\\
	& - \frac{1}{CW_i}\sum_{k = 0}^{CW_i-1}\sum_{j=0}^{k}{\frac{d x_j^i}{d p_i}}.
	\end{align*}
	
	From the above
	\begin{align*}
	-\frac{d J_i}{d p_i} = \frac{\sum\limits_{k = 0}^{CW_i-1}\Big(J_i\big(\frac{d x_k^i}{d p_i}(1-p_i)+(1-x_k^i)\big)+\sum\limits_{j=0}^{k-1}{\frac{d x_j^i}{d p_i}}\Big)}{CW_i - \sum_{k = 0}^{CW_i-1}{(1-x_k^i)(1-p_i)}}.
	\end{align*}
	
	Note that $d x_j^i/d p_i = j(x_j^i-x_{j-1}^i)/p_i$. Furthermore, it holds $J_i \geq 1/p_i$. This can be seen as follows. For $d_i = 0$ it can be seen from \eqref{eq-ji2} that $J_i$ is equal to the average of a geometric random variable of probability $p_i$. By rewriting \eqref{eq-ji} as follows, it can also be seen that $J_i$ becomes larger as $d_i$ increases (indeed, when $d_i$ increases, the terms $x_j^i$ decrease for all $j$, hence the numerator increases and the denominator decreases):
	\begin{align}\label{eq-jirewritten}
	J_i = \frac{\sum_{k = 0}^{CW_i-1}{\Big((k+1) - \sum_{j = 0}^{k}{x_j^i}\Big)}}{CW_i - \sum_{k = 0}^{CW_i-1}{(1-x_k^i)(1-p_i)}}.
	\end{align}
	
	Therefore, $J_i = 1/p_i$ for $d_i = 0$ and $J_i > 1/p_i$ for any other $d_i$. From this,
	\footnotesize
	\begin{align}\label{eq-partialji}
	-\frac{d J_i}{d p_i} \leq \frac{J_i\sum\limits_{k = 0}^{CW_i-1}\Big({\frac{d x_k^i}{d p_i}(1-p_i)+(1-x_k^i)}+\sum\limits_{j = 0}^{k-1}{j(x_j^i-x_{j-1}^i)}\Big)}{CW_i - \sum_{k = 0}^{CW_i-1}{(1-x_k^i)(1-p_i)}}.
	\end{align}
	\normalsize
	
	Recall that $1 - x_k^i$ is the probability of the event that there have been $d_i$ or fewer transmissions in $k$ time slots, from which
	\begin{align*}
	1 - x_k^i \geq \frac{k!}{(k-d_i)!d_i!}p_i^{d_i}(1-p_i)^{k-d_i}.
	\end{align*}
	From the above and~\eqref{eq-partialkx}, we have
	\begin{align*}
	1 - x_k^i \geq \left(\frac{1-p_i}{k-d_i}\right)\frac{d x_k^i}{d p_i} \geq \frac{1-p_i}{k}\frac{d x_k^i}{d p_i}
	\end{align*}
	hence, $(1-p_i)d x_k^i/d p_i \leq k(1 - x_k^i)$.
	
	Combining the above with \eqref{eq-partialji} yields
	\small
	\begin{align*}
	-\frac{d J_i}{d p_i} \leq \frac{J_i\sum\limits_{k = 0}^{CW_i-1}\Big({(k+1)(1-x_k^i)}+\sum\limits_{j = 0}^{k-1}{j(x_j^i-x_{j-1}^i)}\Big)}{CW_i - \sum_{k = 0}^{CW_i-1}{(1-x_k^i)(1-p_i)}},
	\end{align*}
	\normalsize
	and combining the above with~\eqref{eq-ji} we obtain $-d J_i/d p_i \leq J_i^2$, which proves the lemma due to $j_i = 1/J_i$.
\end{IEEEproof}

The following lemmas relate to the stationary regime of system~\eqref{eq-dynsys} (see Sections~\ref{sec:analysis_stat},~\ref{sec:uniqueness}).
Let $\Phi(p_e) = \prod_{k = 0}^{m-1}{(1-\tau_k(p_e))^{\hat{n}_k(p_e)}}.$
The following lemmas examine the function $\Phi(p_e),$ where each $\hat{n}_k(p_e)$ is a function of $\beta_i(p_e),\,p_i(p_e),\,\tau_i(p_e),$ $0\le i \le m-1$, as given by~\eqref{eq-ni}. 
\begin{lm}
	Let $\Phi(p_e) = \prod_{k = 0}^{m-1}{(1-\tau_k(p_e))^{\hat{n}_k(p_e)}}.$ Then, if~\eqref{eq-cond} is satisfied, $\partial \Phi/\partial \beta_j > 0,$ for any $0 \le j < m-1$.
	\label{lemma:betai_inc}
\end{lm}
\begin{IEEEproof}
	We consider the expression $\prod_{k= 0}^{m-1}{(1-\tau_k)^{\hat{n}_k}}$ as a function of $\tau_i,\,p_i$ and $\beta_i$, where $\hat{n}_i$ is computed as a function of $\tau_i,\,\beta_i$ and $p_i$ from (\ref{eq-ni}).
	We show that if we increase $\beta_j$ to $\beta_j^*$ for a given $j$, and leave the remaining $\tau_i,\,p_i$ and $\beta_i$ values fixed, then $\prod_{k= 0}^{m-1}{(1-\tau_k)^{\hat{n}_k}}$ increases.
	First, we have $\partial \Phi/\partial \beta_{m-1} = 0$, because $\Phi(p_e)$ does not depend on $\beta_{m-1}$. We next study the cases with $0 \le j < m-1.$
	
	From~\eqref{eq-ki}, it can be seen that the new $K_i$ values resulting from $\beta_j^*$, denoted by $K_i^*$, satisfy the following.
	If $j=0$, then $K_1^* > K_1$ and $K_i^*=K_i,\,i>1$ by~\eqref{eq-ki}. Thus, $\hat{n}_0^* < \hat{n}_0$ and $\hat{n}_i^* > \hat{n}_i, \, 0<i\le m-1.$
	%\footnote{Here and in following inequalities we use the fact that the function $y(x) = ax/(b+c x)$ with $a,b,c\in\mathbb{R^+}$ is increasing. Replacing $x$ with the $K_j$ and $a,b,c$ with products of $K_i$ that remain constant yields the corresponding relations.}
	%	If $j=m-1,$ then $K_i^*=K_i\text{ and }\hat{n}_i^*=\hat{n}_i,\,0\le i \le m-1 .$
	For $1 \le j \le m-2,$ we have 
	$\prod_{n = 1}^{i}{K_n^*} = \prod_{n = 1}^{i}{K_n}, \ i < j $ and $\prod_{n = 1}^{j}{K_n^*} < \prod_{n = 1}^{j}{K_n}$.
	%\begin{equation}
	%\prod_{n = 1}^{i}{K_n^*} < \prod_{n = 1}^{i}{K_n}, \ i = j
	%\end{equation}
	We also  have $\prod_{n = 1}^{i}{K_n^*} > \prod_{n = 1}^{i}{K_n}, \ i > j,$ since 
	\begin{align}
	& \frac{\prod_{n = 1}^{i}{K_n^*}}{\prod_{n = 1}^{i}{K_n}} = \frac{\frac{\tau_jp_j + \beta_j^*}{\tau_j + \beta_j^*}}{\frac{\tau_jp_j + \beta_j}{\tau_j + \beta_j}},\,
	\frac{\partial}{\partial \beta_j}\left(\frac{\tau_jp_j + \beta_j}{\tau_j + \beta_j}\right) =  \frac{\tau_j (1-p_j)}{(\tau_j+\beta_j)^2} > 0.\nonumber
	\end{align}
	
	Let $\sigma = \sum_{i=1}^{m-1}\prod_{n = 1}^{i}{K_n}.$ We now show that $\sigma^* >\sigma,$ i.e., $\partial \sigma/{\partial \beta_j}>0.$ 
	For $j=m-2,$ $\partial \sigma/{\partial \beta_j}>0$ if and only if $\tau_{m-2}(1-p_{m-2})-\tau_{m-1}(1-p_{m-1})>0,$ which holds by~\eqref{eq-cond}\footnote{According to~\eqref{eq-cond}, $\tau_i$ decreases with $i$. From $p_i = 1 - p_e/(1-\tau_i)$, we have that $p_i$ increases with~$i$.}.

	For $j<m-2,$ we have
	\begin{equation}\label{eq-psigma}
	\footnotesize
	\frac{\partial \sigma}{\partial \beta_j} = \prod_{l = 1}^{j-1}\frac{K_l}{\tau_j+\beta_j}{\left(-K_j +  \frac{K_j \tau_j (1-p_j)}{\tau_{j+1}+\beta_{j+1}} \left(1+ \sum_{i=j+2}^{m-1}\prod_{n = j+2}^{i}{K_n}\right)\right) }.
	\end{equation}
	We prove $\partial \sigma/{\partial \beta_j}>0$ by induction. We first show that $\partial \sigma/{\partial \beta_j}>0$ for $j=m-3,$ and then prove that if this holds for $j=k$, then it also holds for $j=k-1.$
	From~\eqref{eq-psigma}, it can be seen that we need to show that:
	\begin{equation}
	\tau_j(1-p_j)\left(1+\sum_{i=j+2}^{m-1}\prod_{n = j+2}^{i}{K_n}\right)-\tau_{i+1}-\beta_{i+1}>0.
	\label{eq-der_sum_k}
	\end{equation}
	For $j=m-3$, the above holds
	% \begin{align*}
	%  &\tau_{m-3}(1-p_{m-3})(1+K_{m-1})-\tau_{m-2}-\beta_{m-2}\\
	%  &>\tau_{m-3}(1-p_{m-3})+\tau_{m-2}p_{m-2}-\tau_{m-2}>0,
	%  \end{align*}
	because of~\eqref{eq-cond}.
	Now, assume that $\sigma^* >\sigma$ for $j=k.$ Then, we show that $\sigma^* >\sigma$ holds also for $j=k-1.$ 
	% Since $\sigma^* >\sigma$ for $j=k,$~\eqref{eq-der_sum_k} yields
	% \begin{equation}       
	% 1+\sum_{i=k+2}^{m-1}\prod_{n = k+2}^{i}{K_n}>\frac{\tau_{k+1}+\beta_{k+1}}{\tau_{k}(1-p_{k})}.
	% \label{eq-finalk}
	% \end{equation} 
	Let us evaluate~\eqref{eq-der_sum_k} at $j=k-1.$ By using~\eqref{eq-der_sum_k} as true for $j=k$, we have
	\begin{align*}
	& \tau_{k-1}(1-p_{k-1})\left(1+\sum_{i=k+1}^{m-1}\prod_{n = k+1}^{i}{K_n}\right) > \\
	& \quad\quad\quad \tau_{k-1}(1-p_{k-1})\left(1+K_{k+1}\frac{\tau_{k+1}+\beta_{k+1}}{\tau_{k}(1-p_{k})}\right).
	\end{align*}
	Given~\eqref{eq-ki} for $K_{k+1}$ and~\eqref{eq-cond}, the right-hand side of the above inequality yields
	\begin{align*}
	\tau_{k-1}(1-p_{k-1})\left(1+\frac{\tau_{k} p_{k} + \beta_{k}}{\tau_{k}(1-p_{k})}\right) & > \tau_{k}(1-p_{k}) + \tau_{k} p_{k} + \beta_{k}\\
	& = \tau_{k}+\beta_{k}.
	\end{align*}
	Thus, we have proven that if $\sigma^* >\sigma$ for $j=k,$ then the same holds for $j=k-1.$ 
	Given $\sigma^* >\sigma$ it follows that $\hat{n}_{i}^* < \hat{n}_{i}$ for $i \le j .$ Also, since $\hat{n}_{i}^* < \hat{n}_{i},\, i \le j $ and $\sum_{k}{\hat{n}_k^*} = \sum_{k}{\hat{n}_k} = N,$ there must be some $l>j$
	for which $\hat{n}_{l}^* > \hat{n}_{l}.$ Since for $i\ge l+1,$ we have $\hat{n}_i = K_i \hat{n}_{i-1}$ with $K_i^*=K_i,$ it holds that $\hat{n}_{i}^* > \hat{n}_{i},\, i > l .$ Thus,
	\begin{align*}
	\frac{\prod_{k= 0}^{m-1}{(1-\tau_k)^{\hat{n}_k^*}}}{\prod_{k= 0}^{m-1}{(1-\tau_k)^{\hat{n}_k}}} & = \prod_{k < l}{(1-\tau_k)^{\hat{n}_k^*-\hat{n}_k}}\prod_{k \geq l}{(1-\tau_k)^{\hat{n}_k^*-\hat{n}_k}} \nonumber\\
	&> (1-\tau_{l})^{\sum_{k<l}{\hat{n}_k^*-\hat{n}_k}}(1-\tau_{l})^{\sum_{k \geq l}{\hat{n}_k^*-\hat{n}_k}},
	\end{align*}
	because of~\eqref{eq-cond}.
	As $\sum_{k}{\hat{n}_k^*} = \sum_{k}{\hat{n}_k} = N$, the above is larger than 1, which proves the lemma.
\end{IEEEproof}

\begin{lm}
	Let $\Phi(p_e) = \prod_{k = 0}^{m-1}{(1-\tau_k(p_e))^{\hat{n}_k(p_e)}}.$ Then, if~\eqref{eq-cond} is satisfied, $\partial \Phi/\partial p_j > 0,$ for any $0 \le j \le m-1$.
	\label{lemma:pi_inc}
\end{lm}
\begin{IEEEproof}
	The proof is similar to the one of Lemma~\ref{lemma:betai_inc}.
	It can be easily seen from~\eqref{eq-ki} that if $p_j$ increases to $p_j^*$, we have
	\begin{equation*}
	\prod_{n = 1}^{i}{K_n^*} = \prod_{n = 1}^{i}{K_n}, \ i \leq j
	\text{ and }
	\prod_{n = 1}^{i}{K_n^*} > \prod_{n = 1}^{i}{K_n}, \ i > j.
	\end{equation*}
	Note that the above holds for $0\le j \le m-2.$ 
	For $j= m-1 $ we have 
	\begin{equation*}
	\prod_{n = 1}^{i}{K_n^*} = \prod_{n = 1}^{i}{K_n}, \ i < m-1
	\text{, }
	\prod_{n = 1}^{i}{K_n^*} > \prod_{n = 1}^{i}{K_n}, \ i = m-1.
	\end{equation*}
	
	Thus, as $\sigma^* >\sigma$ (with $\sigma = \sum_{i=1}^{m-1}\prod_{n = 1}^{i}{K_n}$) also holds here, it is $\hat{n}_i^* < \hat{n}_i$ for $i \leq j$ and $\hat{n}_i^* > \hat{n}_i$ for $i > j,$ with $0\le j \le m-2.$  For $j= m-1 $ we have $\hat{n}_i^* < \hat{n}_i$ for $i  < m-1$ and $\hat{n}_i^* > \hat{n}_i$ for $i = m-1$.
	Then, following the same reasoning as for the previous lemma, it can be seen that $\prod_{k= 0}^{m-1}{(1-\tau_k)^{\hat{n}_k^*}} > \prod_{k= 0}^{m-1}{(1-\tau_k)^{\hat{n}_k}}$, which proves the lemma.
\end{IEEEproof}

\begin{lm}
	If~\eqref{eq-cond} is satisfied, then $\partial \Phi/\partial \tau_j < 0,$ for any $0 \le j \le m-1$.
	\label{lemma:taui_inc}
\end{lm}
\begin{IEEEproof}
	When $\tau_j$ increases to $\tau_j^*$, $\prod_{n = 1}^{i}{K_n^*} = \prod_{n = 1}^{i}{K_n} $ for $i < j,$ and $\prod_{n = 1}^{i}{K_n^*} < \prod_{n = 1}^{i}{K_n} $ for $i = j$. 
	For $i>j$ we have
	\begin{equation*}
	\prod_{n = 1}^{i}{K_n^*} = \prod_{n = 1}^{i}{K_n}\frac{\frac{\tau_j^*p_j + \beta_j}{\tau_j^* + \beta_j}}{\frac{\tau_jp_j + \beta_j}{\tau_j + \beta_j}},\text{ and }\frac{\partial}{\partial \tau_j}\left(\frac{\tau_jp_j + \beta_j}{\tau_j + \beta_j}\right)  < 0.
	\end{equation*}
	Thus, $\prod_{n = 1}^{i}{K_n^*} < \prod_{n = 1}^{i}{K_n}$ for $i\ge j$. 
	These yield $\sigma^* < \sigma$ (with $\sigma = \sum_{i=1}^{m-1}\prod_{n = 1}^{i}{K_n}$). From the above, $\hat{n}_i^* > \hat{n}_i$ for $i < j$ and $\hat{n}_j^* < \hat{n}_j,$ because  $K_j^* < K_j.$  
	%	 \begin{equation}
	%	 \sum_{i=1}^{m-1}\prod_{n = 1}^{i}{K_n^*} < \sum_{i=1}^{m-1}\prod_{n = 1}^{i}{K_n}.
	%	 \label{eq-sum_ki2}
	%	 \end{equation}
	Hence, by using similar arguments as in the two previous lemmas, it follows that $\prod_{k \neq j}{(1-\tau_k)^{\hat{n}_k}}$ decreases. If we show that $(1-\tau_j)^{\hat{n}_j}$ also decreases, the lemma will be proven. Note that
	
	%	We distinguish two cases for $i>j.$ Suppose that  $\hat{n}_i^* < \hat{n}_i$ for $i > j.$ Then, following the same reasoning as for Lemma \ref{lemma:betai_inc}, it can be seen that $\prod_{k \neq j}{(1-\tau_k)^{\hat{n}_k}}$ decreases. 
	% Suppose now that $\hat{n}_i^* > \hat{n}_i$ for $i > j.$ Then, $\prod_{k \neq j}{(1-\tau_k)^{\hat{n}_k}}$ decreases because it is a product of the decreasing functions $(1-\tau_k)^{\hat{n}_k}$ with respect to $\hat{n}_k.$ 
	\begin{equation}
	\frac{\partial (1-\tau_j)^{\hat{n}_j}}{\partial \tau_j} = - \hat{n}_j(1-\tau_j)^{\hat{n}_j-1} + \ln(1-\tau_j)\frac{\partial \hat{n}_j}{\partial \tau_j}(1-\tau_j)^{\hat{n}_j}.
	\label{eq-ptj}
	\end{equation}
	
	Computing the partial derivative of $\hat{n}_j,$ we have
	\begin{equation}
	\frac{\partial \hat{n}_j}{\partial \tau_j} =  - \frac{\hat{n}_j}{\tau_j + \beta_j} + \frac{\partial \hat{n}_{j-1}}{\partial \tau_{j}}K_j \geq - \frac{\hat{n}_j}{\tau_j + \beta_j} \geq - \frac{\hat{n}_j}{\tau_j},
	\label{eq-pnj}
	\end{equation}
	because $\partial \hat{n}_{j-1}/\partial \tau_{j} >0$ from above. This inequality holds for any $0 < j \le m-1.$ By taking into account that $n_0 = N/\sigma,$ it is easy to see that it also holds for $j=0.$ Note that, in this case, $\partial \sigma / \partial \tau_0 > 0.$ For this case, we have 
	\begin{equation*}
	\frac{\partial \hat{n}_0}{\partial \tau_0} =  - \frac{\hat{n}_0}{\sigma} \frac{(\sigma -1) p_0}{\tau_0 p_0 + \beta_0} \geq - \frac{  \hat{n}_0 p_0}{\tau_0 p_0 + \beta_0} \geq - \frac{\hat{n}_0}{\tau_0}.
	\end{equation*}
	
	Combining~\eqref{eq-ptj} and~\eqref{eq-pnj} yields
	\begin{equation*}
	\frac{\partial (1-\tau_j)^{\hat{n}_j}}{\partial \tau_j} \leq \frac{\hat{n}_j(1-\tau_j)^{\hat{n}_j}}{\tau_j} \left(-\frac{\tau_j}{1-\tau_j} - \ln(1-\tau_j)\right).
	\end{equation*}
	
	As $-x/(1-x) < \ln(1-x)$, the above is smaller than 0, which proves the lemma.
\end{IEEEproof}

{\color{black}
	%To prove the global asymptotic stability of the system~\eqref{ode} for $m = 3$, we use the following lemmas.
	
	The following lemma and proof relate to the asymptotic continuous-time system (see Section~\ref{sec:ode}). Recall that $y_i$ denotes the proportion of stations at backoff stage $i,$ and that $\tau_i$ and $\beta_i$ are now functions of $\rho = 1-e^{-\sum_{k=0}^{m-1} y_k \tau_k(\rho)}$ that is the collision probability and is independent of the state of the system compared to all previous Appendix analysis.
	
	\begin{lm}\label{lm-partialgamma} Let $\gamma = \tau_0 y_0 + \tau_1 y_1 + \tau_2 (1-y_0-y_1)$. If~\eqref{eq-cond} is satisfied, it holds that $0 < \partial \gamma/\partial y_i \leq \tau_i$.
	\end{lm}
	\begin{IEEEproof}
		The partial derivative can be expressed as
		\begin{align}\label{partialgamma}
		\frac{\partial \gamma}{\partial y_i} = &\, \frac{\partial}{\partial y_i}\left( \tau_0 y_0 + \tau_1 y_1 + \tau_2 (1-y_0-y_1) \right) \nonumber\\
		= &\ \tau_i - \tau_2 + y_0 \frac{\partial \tau_0}{\partial \gamma}\frac{\partial \gamma}{\partial y_i} + y_1 \frac{\partial \tau_1}{\partial \gamma}\frac{\partial \gamma}{\partial y_i}  \nonumber\\
		& + \ (1- y_0 - y_1) \frac{\partial \tau_2}{\partial \gamma}\frac{\partial \gamma}{\partial y_i},
		\end{align}
		from which
		\begin{equation}\label{partialgamma2}
		\frac{\partial \gamma}{\partial y_i} = \frac{\tau_i - \tau_2}{1- y_0 \frac{\partial \tau_0}{\partial \gamma} - y_1 \frac{\partial \tau_1}{\partial \gamma} - (1- y_0 - y_1) \frac{\partial \tau_2}{\partial \gamma}}.
		\end{equation}
		
		From Lemma {\ref{lemma:tau_bci}}, we have that $\tau_i$ is a decreasing function of $\rho = 1 - e^{-\gamma}$, hence $\partial \tau_i/\partial \gamma \leq 0$. Furthermore, due to~\eqref{eq-cond}, we have $\tau_i - \tau_2 > 0$ for $i=0,\,1.$ Combining this with~\eqref{partialgamma2} yields $\partial \gamma/\partial y_i > 0$. Now, combining \eqref{partialgamma} with $\partial \tau_i/\partial \gamma \leq 0$ and $\partial \gamma/\partial y_i \geq 0$ yields $\partial \gamma/\partial y_i \leq \tau_i$.
	\end{IEEEproof}
	
	\vspace{0,4cm}
	
	\begin{IEEEproof}[Proof of Theorem~\ref{thm:ode3}]
		For $m = 3$, the system \eqref{ode} can be expressed with the following two equations as a function of $y_0$ and $y _1$ (where $y_2 = 1 - y_0 - y_1$):
		\begin{align*}
		\frac{d y_0}{d t} & = \gamma e^{-\gamma} - y_0 (\tau_0 + \beta_0) = f_0 (y_0,y_1)\\
		\frac{d y_1}{d t} & = y_0(\tau_0(1-e^{-\gamma})+\beta_0)-y_1(\tau_1 + \beta_1) = f_1 (y_0,y_1).
		\end{align*}
		where $\gamma = \tau_0 y_0 + \tau_1 y_1 + \tau_2 (1-y_0-y_1)$ (to simplify notation, we have omitted the dependency of $\tau_i$ and $\beta_i$ on $\gamma$ in the above equations). $\tau_i$ and $\beta_i$ depend on $\gamma$ through $1-e^{-\gamma}$ that is the collision probability here. Thus, we will use the same lemmas proven for the collision probability denoted as $p_i$ before. 
		
		According to the Markus-Yamabe theorem,\footnote{\color{black}This theorem was initially stated as a conjecture for any $n$, and was later proved to be true for $n = 2$ (which is the case here) and false for $n>2$~\cite{fessler}.} the above system is {globally asymptotically stable if} the real part of the eigenvalues of the following Jacobian matrix are negative for all points ($y_0$, $y_1$):
		\renewcommand{\arraystretch}{0.6}
		\begin{equation*}
		\left( \begin{array}{cc}
		\dfrac{\partial f_0}{\partial y_0} & \dfrac{\partial f_0}{\partial y_1} \\\\
		\dfrac{\partial f_1}{\partial y_0} & \dfrac{\partial f_1}{\partial y_1}
		\end{array} \right).
		\end{equation*}
		
		The characteristic polynomial of  the matrix is given by
		\begin{equation*}
		\lambda^2 - (c_{00} + c_{11}) \lambda + c_{00} c_{11} - c_{01} c_{10},
		\end{equation*}
		where $c_{ij} = \partial f_i/\partial y_j$. 
		
		According to the Routh-Hurwitz stability criterion, as long as the following two inequalities are satisfied, the real part of the eigenvalues of the above characteristic polynomial are guaranteed to be negative:
		\begin{equation}\label{hurwitz1}
		- (c_{00} + c_{11}) > 0,
		\end{equation}
		\begin{equation}\label{hurwitz2}
		c_{00} c_{11} - c_{01} c_{10} > 0.
		\end{equation}
		
		Sufficient conditions for \eqref{hurwitz1} are ($i$) $c_{00} < 0$ and ($ii$) $c_{11} < 0$, and sufficient conditions for \eqref{hurwitz2} are: ($iii$) $-c_{00} > c_{10}$, ($iv$) $-c_{11} > c_{01}$ and ($v$) $c_{10}$ and $c_{01}$  are never negative simultaneously. In the following, we show that each of these five conditions is satisfied.
		
		\textbf{Condition ($i$): $c_{00} < 0$}.
		\begin{align}\label{eq-c00}
		c_{00} = \frac{\partial f_0}{d y_0} = & \ (1-\gamma)e^{-\gamma} \frac{\partial \gamma}{\partial y_0} - (\tau_0+\beta_0)  \\
		& - y_0 \frac{\partial (\tau_0+\beta_0)}{\partial (1-e^{-\gamma})}\frac{\partial (1-e^{-\gamma})}{\partial \gamma}\frac{\partial \gamma}{\partial y_0}.\nonumber
		\end{align}

		Removing some negative terms from the above equation and considering that $\partial (\tau_0+\beta_0)/\partial (1-e^{-\gamma}) > 0,$ $\partial \gamma/{\partial y_0} > 0$ (see Lemma \ref{lm-partialgamma}) and $(1-\gamma)e^{-\gamma} < 1$, we obtain the following upper bound on $c_{00}$:
		\begin{equation*}
		c_{00} < \frac{\partial \gamma}{\partial y_0} - {\tau_0}.
		\end{equation*}
		
		From Lemma \ref{lm-partialgamma}, the left-hand side of the above is smaller than or equal to 0.

		\textbf{Condition ($ii$): $c_{11} < 0$}.
		\begin{align*}
		c_{11} = & \ y_0 \frac{\partial (\tau_0(1-e ^{-\gamma})+\beta_0)}{\partial (1-e^{-\gamma})}\frac{\partial (1-e^{-\gamma})}{\partial \gamma}\frac{\partial \gamma}{\partial y_1} \nonumber\\
		& -(\tau_1+\beta_1) - y_1 \frac{\partial (\tau_1+\beta_1)}{\partial (1-e^{-\gamma})}\frac{\partial (1-e^{-\gamma})}{\partial \gamma}\frac{\partial \gamma}{\partial y_1}  \nonumber\\
		\leq \ & y_0 \frac{\partial (\tau_0(1-e ^{-\gamma})+\beta_0)}{\partial (1-e^{-\gamma})}\frac{\partial (1-e^{-\gamma})}{\partial \gamma}\frac{\partial \gamma}{\partial y_1} - \tau_1.
		\end{align*}
		
		From Lemma \ref{lm-ji} and by replacing $p_i$ with $1-e^{-\gamma}$, we have 
		$$
		\frac{	\partial (\tau_0(1-e ^{-\gamma})+\beta_0)}{\partial (1-e^{-\gamma})} \le 1.
		$$ 
		
		By using the above inequality, Lemma \ref{lm-partialgamma} and $e^{-\gamma} < 1$, we obtain the following upper bound
		\begin{equation*}
		c_{11} < y_0 \tau_1 - \tau_1 \leq 0.
		\end{equation*}
		
		\textbf{Condition ($iii$): $- c_{00} > c_{10}$}.
		
		We have
		\begin{align*}
		c_{10} = & \ \tau_0 + \beta_0 + y_0 \frac{\partial (\tau_0+\beta_0)}{\partial (1-e^{-\gamma})}\frac{\partial (1-e^{-\gamma})}{\partial \gamma}\frac{\partial \gamma}{\partial y_0} - e^{-\gamma} \tau_0 \nonumber\\
		&- \ y_0 \frac{\partial (e^{-\gamma} \tau_0)}{\partial (1-e^{-\gamma})}\frac{\partial (1-e^{-\gamma})}{\partial \gamma}\frac{\partial \gamma}{\partial y_0}  \nonumber\\
		&- \ y_1 \frac{\partial (\tau_1+\beta_1)}{\partial (1-e^{-\gamma})}\frac{\partial (1-e^{-\gamma})}{\partial \gamma}\frac{\partial \gamma}{\partial y_0}.
		\end{align*}
		
		Comparing the above equation with \eqref{eq-c00}, it can be seen that it holds $- c_{00} > c_{10}$ as long as
		\begin{align*}
		-(1-\gamma)e^{-\gamma} \frac{\partial \gamma}{\partial y_0} > & -e^{-\gamma} \tau_0 - \ y_0 \frac{\partial (e^{-\gamma} \tau_0)}{\partial (1-e^{-\gamma})}\frac{\partial (1-e^{-\gamma})}{\partial \gamma}\frac{\partial \gamma}{\partial y_0} \nonumber\\
		&  - \ y_1 \frac{\partial (\tau_1+\beta_1)}{\partial (1-e^{-\gamma})}\frac{\partial (1-e^{-\gamma})}{\partial \gamma}\frac{\partial \gamma}{\partial y_0}.
		\end{align*}
		
		Removing some terms and rearranging others, we obtain the following sufficient condition for the above
		\begin{align*}
		e^{-\gamma} \left(\tau_0 - \frac{\partial \gamma}{\partial y_0}\right) & > -\gamma e^{-\gamma} \frac{\partial \gamma}{\partial y_0}- y_0 \frac{\partial (e^{-\gamma} \tau_0)}{\partial \gamma}\frac{\partial \gamma}{\partial y_0}.
		\end{align*}
		
		Following a similar reasoning to Lemma \ref{lm-partialgamma}, it can be seen that $\partial \gamma/\partial y_0 < \tau_0 + y_0 \partial \tau_0/\partial y_0$, from which we obtain the following sufficient condition
		\begin{align*}
		-e^{-\gamma} y_0 \frac{\partial \tau_0}{\partial y_0} & > -\gamma e^{-\gamma} \frac{\partial \gamma}{\partial y_0}- y_0 \frac{\partial (e^{-\gamma} \tau_0)}{\partial \gamma}\frac{\partial \gamma}{\partial y_0}.
		\end{align*}
		Taking into account that $\frac{\partial \tau_0}{\partial y_0} = \frac{\partial \tau_0}{\partial \gamma}\frac{\partial \gamma}{\partial y_0}$, this is equivalent to 
		\begin{align*}
		-e^{-\gamma} y_0 \frac{\partial \tau_0}{\partial \gamma} & > -\gamma e^{-\gamma} - y_0 \frac{\partial (e^{-\gamma} \tau_0)}{\partial  \gamma}.
		\end{align*}
		
		Operating on the above we obtain
		\begin{align*}
		-e^{-\gamma} y_0 \frac{\partial \tau_0}{\partial \gamma} & > -\gamma e^{-\gamma} - y_0 e^{-\gamma} \frac{\partial \tau_0}{\partial  \gamma} + y_0 e^{-\gamma} \tau_0.
		\end{align*}
		
		The above is equivalent to
		\begin{align*}
		0 > - e^{-\gamma} (\gamma - y_0 \tau_0),
		\end{align*}
		which is true given that $\gamma > y_0 \tau_0$ for $y_0 < 1$ (for the case $y_0 = 1$ it can be seen that this condition also holds).
		
		\textbf{Condition ($iv$): $- c_{11} > c_{01}$}.
		
		A sufficient condition for this case is given by
		\begin{align}\label{eq-cond4}
		y_0\frac{\partial (e^{-\gamma} \tau_0)}{\partial \gamma} \frac{\partial \gamma}{\partial y_1} + (\tau_1+\beta_1) + y_1 \frac{\partial (\tau_1+\beta_1)}{\partial \gamma} \frac{\partial \gamma}{\partial y_1}  \nonumber\\
		> (1-\gamma)e^{-\gamma}\frac{\partial \gamma}{\partial y_1}.
		\end{align}
		The left-hand side of~\eqref{eq-cond4} is lower bounded by
		\begin{align*}
		& \tau_1 - y_0 e^{-\gamma} \tau_0 \frac{\partial \gamma}{\partial y_1} + y_0 e^{-\gamma} \frac{\partial \tau_0}{\partial \gamma}\frac{\partial \gamma}{\partial y_1}.
		\end{align*}
		
		Given $\frac{\partial \tau_0}{\partial \gamma}\frac{\partial \gamma}{\partial y_1} = \frac{\partial \tau_0}{\partial y_1} $ and $\frac{\partial \gamma}{\partial y_1} < \tau_1$, the above is lower bounded by
		\begin{align*}
		& \tau_1 - y_0 e^{-\gamma} \tau_0 \tau_1 +  y_0 e^{-\gamma} \frac{\partial \tau_0}{\partial y_1} = \nonumber\\
		& \tau_1 e^{-\gamma} ( 1- y_0 \tau_0) + \tau_1 (1-e^{-\gamma})  + \frac{\partial \tau_0}{\partial y_1} y_0 e^{-\gamma}.
		\end{align*}
		
		From $1-e^{-x} > xe^{-x}$, the above is lower bounded by
		\begin{align}\label{eq-lowerbound}
		\tau_1 e^{-\gamma} ( 1- y_0 \tau_0) + \tau_1 \gamma e^{-\gamma} + \frac{\partial \tau_0}{\partial y_1} y_0 e^{-\gamma}.
		\end{align}
		
		It can be seen that $0 > \partial \tau_0/\partial \gamma > -1$. From the proof of Corollary \ref{cor:bi-der}, $\frac{\partial B_0}{\partial (1-e^{-\gamma})} \leq \frac{B_0^2}{e^{-\gamma}}$, where $B_0 = \frac{1}{\tau_0}-1$. Thus, 
		\begin{align}
		\frac{\partial \tau_0}{\partial \gamma} = -\frac{1}{(B_0+1)^2}\frac{\partial B_0}{\partial (1-e^{-\gamma})}\frac{\partial (1-e^{-\gamma})}{\partial \gamma} > -1.
		\end{align}
		
		Combining this with Lemma \ref{lm-partialgamma} yields $|\partial \tau_0/\partial y_1| < \tau_1 $, from which $\tau_1 \gamma e^{-\gamma} > - \gamma e^{-\gamma} y_0 \frac{\partial \tau_0}{\partial y_1}$. Combining this with \eqref{eq-lowerbound} yields the following lower bound for the LHS of \eqref{eq-cond4}
		\begin{align}\label{eq-lowerbound2}
		\tau_1 e^{-\gamma} ( 1- y_0 \tau_0) +  \frac{\partial \tau_0}{\partial y_1} y_0 e^{-\gamma} (1-\gamma).
		\end{align}
		
		From $\partial \gamma/\partial y_1 < \tau_1 + y_0 \partial \tau_0/\partial y_1$, the right-hand side of \eqref{eq-cond4} is upper bounded by 
		\begin{equation}\label{eq-upperbound}
		(1-\gamma)e^{-\gamma}\tau_1 + (1-\gamma)e^{-\gamma} y_0 \frac{\partial \tau_0}{\partial y_1}.
		\end{equation}
		
		Comparing \eqref{eq-lowerbound2} with \eqref{eq-upperbound}, it can be seen that that in \eqref{eq-lowerbound2} the positive term is larger and the negative terms are equal, which implies that \eqref{eq-cond4} is satisfied.
		
		\textbf{Condition ($v$): either $c_{10} \geq 0$ or $c_{01} \geq 0$}.
		
		To show that $c_{10}$ and $c_{01}$ are never negative simultaneously, we prove that when $y_0 < 1 - \gamma$ it holds $c_{01} \geq 0$, and otherwise $c_{10} \geq 0$. 
		Let us start with $c_{01}$,
		\begin{align*}
		c_{01}   = & \ (1-\gamma)e^{-\gamma} \frac{\partial \gamma}{\partial y_1} - y_0 \frac{\partial (\tau_0+\beta_0)}{\partial (1-e^{-\gamma})}\frac{\partial (1-e^{-\gamma})}{\partial \gamma}\frac{\partial \gamma}{\partial y_1} \nonumber\\
		= & \ (1-\gamma)e^{-\gamma} \frac{\partial \gamma}{\partial y_1} - y_0 \frac{\partial (\tau_0(1-e^{-\gamma})+\beta_0)}{\partial (1-e^{-\gamma})}e^{-\gamma}\frac{\partial \gamma}{\partial y_1} \nonumber\\
		& \ - y_0 \frac{\partial (\tau_0 e^{-\gamma})}{\partial (1-e^{-\gamma})}e^{-\gamma}\frac{\partial \gamma}{\partial y_1} .
		\end{align*}
		
		Given that $\tau_0$ is a decreasing function of $1-e^{-\gamma}$ (Lemma \ref{lemma:tau_bci}), the last term of the above expression is positive, hence
		\begin{align*}
		c_{01} \geq (1-\gamma)e^{-\gamma} \frac{\partial \gamma}{\partial y_1} - y_0 \frac{\partial (\tau_0(1-e^{-\gamma})+\beta_0)}{\partial (1-e^{-\gamma})}e^{-\gamma}\frac{\partial \gamma}{\partial y_1}.
		\end{align*}
		
		By combining the above with Lemma \ref{lm-ji}, we obtain
		\begin{align*}
		c_{01} \geq \left((1-\gamma)e^{-\gamma} -y_0 e^{-\gamma}\right)\frac{\partial \gamma}{\partial y_1},
		\end{align*}
		which is larger than or equal to 0 as long as $y_0 < 1 - \gamma$.
		
		We now look at $c_{10}$ when $y_0 > 1 - \gamma$,
		\begin{align*}
		c_{10} = & \ \tau_0(1-e^{-\gamma}) + \beta_0 + y_0 \frac{\partial (\tau_0(1-e^{-\gamma})+\beta_0)}{\partial (1-e^{-\gamma})} \\\nonumber
		& \cdot\frac{\partial (1-e^{-\gamma})}{\partial \gamma}\frac{\partial \gamma}{\partial y_0}- y_1 \frac{\partial (\tau_1+\beta_1)}{\partial (1-e^{-\gamma})}\frac{\partial (1-e^{-\gamma})}{\partial \gamma}\frac{\partial \gamma}{\partial y_0}  \\\nonumber
		= & \ \tau_0(1-e^{-\gamma}) + \beta_0 + y_0 \frac{\partial (\tau_0(1-e^{-\gamma})+\beta_0)}{\partial (1-e^{-\gamma})} \frac{\partial (1-e^{-\gamma})}{\partial \gamma}\\\nonumber
		& \cdot\frac{\partial \gamma}{\partial y_0}- y_1 \frac{\partial (\tau_1(1-e^{-\gamma})+\beta_1)}{\partial (1-e^{-\gamma})}\frac{\partial (1-e^{-\gamma})}{\partial \gamma}\frac{\partial \gamma}{\partial y_0}  \\\nonumber
		& - \ y_1 \frac{\partial \tau_1 e^{-\gamma}}{\partial (1-e^{-\gamma})}\frac{\partial (1-e^{-\gamma})}{\partial \gamma}\frac{\partial \gamma}{\partial y_0} \geq \tau_0(1-e^{-\gamma}) \\
		& - \ y_1 \frac{\partial (\tau_1(1-e^{-\gamma})+\beta_1)}{\partial (1-e^{-\gamma})}\frac{\partial (1-e^{-\gamma})}{\partial \gamma}\frac{\partial \gamma}{\partial y_0}.
		\end{align*}
		
		From Lemma \ref{lm-ji} we obtain the following inequality
		\begin{align*}
		c_{10} \geq \tau_0(1-e^{-\gamma}) - y_1 e^{-\gamma} \tau_0.
		\end{align*}
		
		Given $y_0 > 1 - \gamma$, we have $y_1 \leq 1 - y_0 < \gamma$. Substituting this into the above equation yields
		\begin{align*}
		c_{10} \geq \tau_0(1-e^{-\gamma}) - \tau_0 \gamma e^{-\gamma},
		\end{align*}
		which is larger than 0 given that $1-e^{-x} > xe^{-x}$.
	\end{IEEEproof}
}

%\section*{Acknowledgements}
%{\color{black}The authors thank the editor and reviewers for their valuable feedback, which has been very helpful in improving the paper.}

\bibliographystyle{IEEEtran}
\bibliography{IEEEabrv,mybib}
\end{document}